\documentclass[preprint,3p,times]{elsarticle}

\usepackage{amssymb}
\usepackage{amsmath}
\usepackage{multirow}

\usepackage{enumitem}
\usepackage{natbib}
\newcommand{\NHT}{non-hierarchical triangles }
\newcommand{\HT}{hierarchical triangles }
\newcommand{\re}[1]{\textcolor{red}{{[Revision]}}}
\usepackage{bm}
\usepackage{caption}
\usepackage{subcaption}
\usepackage{xcolor}
\usepackage{url}
\usepackage{hyperref}
\hypersetup{
    colorlinks=true,
    linkcolor=blue,
    filecolor=magenta,      
    urlcolor=cyan,
    pdftitle={Overleaf Example},
    pdfpagemode=FullScreen,
    }

\journal{Astronomy and Computing}

\begin{document}

\begin{frontmatter}



\title{The influence of spin in black hole triplets} 


\author[label1,label2,label3]{Ariel Chitan} 
\affiliation[label1]{organization={Physics and Astronomy department, University of Western Ontario},
            addressline={1151 Richmond Street}, 
            city={London},
            postcode={N6A 3K7}, 
            state={Ontario},
            country={Canada}}
            
\affiliation[label2]{organization={Graduate Program in Astrophysics, Observatoire de Paris, Universite  PSL, CNRS},
            addressline={5 pl. Jules Janssen}, 
            postcode={92195}, 
            state={Meudon},
            country={France}}
\affiliation[label3]{organization={Physics department, University of the West Indies},
            state={Trinidad},
            country={W.I.}}

\author[label4]{Aleksandr Myll{\"a}ri}

\affiliation[label4]{organization={Abo Akademi University},
            state={Turku/Abo},
            country={Finland}}

\author[label5,label6,label7]{Mauri Valtonen}
\affiliation[label5]{organization={Finnish Centre for Astronomy with ESO (FINCA), University of Turku},
            state={Turku},
            country={Finland}}

\affiliation[label6]{organization={Tuorla Observatory, Department of Physics and Astronomy, University of Turku},
            state={Turku},
            country={Finland}}

\affiliation[label7]{organization={Visitor, Institute of Astronomy, University of Cambridge, Madingley Road},
            state={Cambridge},
            country={England}}

\begin{abstract}
Spin can influence the dynamics of the already chaotic black hole triplet system. We follow this problem in two sets of simulations: first, the Agekian-Anosova region (or region D), and second, using Pythagorean triangles. We use ARCcode, an N-body code that performs numerical integration of orbits. This code includes post-Newtonian corrections, which we include up to the 2.5$^{th}$ order. In set one of our simulations, we fix the masses of the black holes at 10$^{6}$ M$_{\odot}$. Then we run the simulations first without any spin added and after by initialising spin on one of the black holes. We find that after including spin into the system, 12.9\% of the simulations changed outcomes. Either the systems went from having all black holes merging to having a black hole escaping the system, or vice versa. In the second set of simulations, we expanded into Pythagorean triangles as initial positions of black holes, stemming from Burrau’s three-body problem. We varied the masses of the black holes from 10$^{0}$ M$_{\odot}$ - 10$^{12}$ M$_{\odot}$. Black holes in these systems were given spin in normalised units ranging from 0 to $\sim$ 0.95. We find that intermediate mass black holes in the range of 10$^{4}$ M$_{\odot}$-10$^{5}$ M$_{\odot}$, were influenced the most by spin, particularly in their lifetimes. We also find that simulations, initialised as 2D, become 3D.

\end{abstract}





\end{frontmatter}



\section{Introduction}
\label{intro}

The detection of black hole mergers is becoming more and more commonplace in modern astrophysics due to the advanced LIGO and VIRGO systems \citep{LIGO1,LIGO2}.  The mergers of stellar mass black holes (SBH) have dominated earth-based detections, but it is expected that with space-based detectors, like LISA, finding mergers of intermediate mass black holes (IMBH) as well as supermassive black holes (SMBH) should be possible \citep{Rhook2005}. One theory of origin of these mergers are via isolated binaries \citep{Mcmillan2001,Belcyz2002}. A triplet origin, however, is also a viable theory as two of the black holes can be driven to merge due to the gravitational influence of the third \citep{Rodriguez2018}. Triplet SBH systems may form from isolated triplet star systems \citep{Gomez2021}. IMBH's have so far been been hard to detect, but recent advances have been made by \citet{imbh2024}. Triplet IMBH systems may form within globular clusters \citep{Sigurdsson1993,Vitral2021} and the hierarchical merging of galaxies can potentially lead to the formation of triplet SMBH's \citep{Valtonen1996,Hoffman2007,Kollatschny2020,Yadav2021} as most massive galaxies host a SMBH at their core \citep{Kormendy1995}. With such rapid advancements in the field of gravitational wave physics, the study of the dynamics of triplet black hole systems is necessitated. 

Triplet systems, and the wider three-body, is complex and chaotic. Recent strides have been made to develop a solution to the three-body problem, both analytically \citep{Heggie1996,stone2019,Hamers2019,ginat2021} and numerically \citep{Szebehely1967,monaghan1976,Hut1,Hut2,Hut3,Hut4,Hut5,Hut6,Hut7}. Since the three-body problem has too many parameters we fix masses of the black holes in one set of our simulations and keep all simulations to the free fall case so that we can study, more easily, the effect of spin. We are interested in  the configurations proposed by \citet{Burrau1913}, known as Burrau's problem of three bodies. This problem, though popularized by Burrau, was initially brought to his attention by Ernst Meissel \citep{Peetre1995, valtonenetal2016}. In this problem he describes three bodies placed at the vertices of a 3,4,5 Pythagorean triangle with masses reflecting the side of the triangle opposite to each body. Meissel believed the problem to have a periodic solution while Burrau was unable to find, more rigorously, any proof of this. The evolution of such a system was later studied extensively by both \citet{Szebehely1967} and \citet{valtonen1995}. In the work of the former, they used computers to integrate the orbital motion and found that eventually such a triplet system would end in a bound binary and an escaping body. Valtonen et al. studied, from a relativistic approach, the influence of mass variation from 10$^5$ M$_{\odot}$ to 10$^9$ M$_{\odot}$ in five different cases. They found that for smaller mass systems, binary formation and escape was how the system would typically end but for larger masses, mergers would take place instead. 

In \citet{Chitan2021}, some of the authors reanalysed this work and extended the mass ranges from from 10$^0$ M$_{\odot}$ to 10$^{12}$ M$_{\odot}$ for initial configurations of all Pythagorean triangles with hypotenuse $<$100pc. It was found that as the mass increased, the fraction of mergers also increased and the lifetimes of the triplet systems decayed exponentially. Distinctions were also made between triangles that were close to a hierarchical configuration and those that were not, using the definitions proposed by \citet{Anosova1990}. Here \HT are the triangles where there a distinctive inner binary and a third body orbiting from far away is obvious. For \NHT, this configuration is less obvious. In the more hierarchical type triangles in the study, merging dominated even from low mass black holes (10$^0$ M$_{\odot}$ - 10$^3$ M$_{\odot}$) and their evolution was more predictable than that of the more \NHT. The more \NHT, like the 3,4,5, exhibited a more chaotic nature. We also saw that there was a transition from escape dominated dynamics to merger dominated dynamics in such triplet systems as was shown by \citet{valtonen1995}. \citet{boekholt2021} also studied the same 3,4,5 Pythagorean configuration using the code BRUTUS with 2.5$^{th}$ order post-Newtonian (pN) terms and 1pN cross terms. They also replicated the transition from escapes to mergers in such triplet systems.

All black holes in the triplet systems studied were kept as non-rotating in \citet{Chitan2021}. But, we expect black holes to rotate. \citet{Kerr1963} studied mathematically this more realistic case of rotating black holes. Following the no-hair theorem, a fully isolated black hole can be described by only three parameters - its mass, M, and its spin, J \citep{Israel1967,Israel1968,Hawking1972}, and its charge (which we do not consider here). The next step, which we take in this paper, is study how adding spin into the system will influence the evolution of the black hole triplets. 

As gravitational wave physics and the discovery and study of black holes becomes more prominent in modern astrophysics, it is important to consider what the effect of spin will have on black hole dynamics. Idealized Schwarzschild black holes are much `easier' to deal with than their much more realistic Kerr counterparts. However, almost all astrophysical bodies do rotate and the relativistic effects associated may have an impact on the evolution of n-body systems. One of the major effects of having a rotating black hole in the system is relativistic frame-dragging. The spacetime around the rotating black hole would be dragged and therefore affect other orbiting bodies (in this case the other two black holes of our triplet system). From this comes the Lense-Thirring effect whereby the argument of periapsis as well as the longitude of the ascending node precesses due to the frame-dragging effect of a rotating body in the system \citep{Lense1918,Ciufolini2004}. 

Fang et al. in a series of papers looked at hierarchical triplet systems with a rotating SMBH and inner binary comprised of SBH's \citep{Fang1,Fang2,Fang3}. The typical von Zeipel-Kozai-Lidov oscillations imposed in such systems are with respect to non-rotating bodies. Fang et al. compared the evolution of triplet systems in the hierarchical formation with and without SMBH spin in order to compare the effect of the von Zeipel-Kozai-Lidov oscillations to other effects that may be introduced due to the spin of the SMBH. They found that there were significant differences in the evolution of the inner binary black hole system when spin was included - depending on the initial orbital angles, the merger time of the binary could be lengthened or shortened if spin was present \citep{Fang2}. They also conducted numerical simulations to show that the effect of spin could be detectable by the upcoming LISA. This could potentially lead to new ways of observing the spin effect in triplet systems as well as the spin parameters of SMBH's \citep{Fang3}.

\citet{kidder1995} discusses in depth the effect of spin on a coalescing binary in the pN regime. They show that one of the major contributions of these spin terms to the orbital dynamics is that they are not confined to the orbital plane. They then result in the precession of the orbital plane if the spin vector is not aligned perpendicular to the orbital plane. They also demonstrate the spin contribution to energy loss as well as the angular momentum loss in gravitational radiation, where they show that the spin-orbit contribution can be significant. The role of spin in the evolution of such systems can be expected to have a significant effect. This provides the motivation to test this effect via numerical simulations on not binaries but on the more complex but also astrophysically common triplet systems.

In this paper, we extend the work of \citet{Chitan2021} by allowing the most massive black hole in the triplet system to rotate and conducting numerical simulations using ARCcode. The spin vector as well as the masses of the black holes are varied. This is presented in three sections - in \textbf{Section} \ref{sect:2} we will focus on the method and dataset. In \textbf{Section} \ref{sect: Section 3}, the results for the effect of increasing spin vector and increasing mass will be presented and discussed. 

\section{Setup}\label{sect:2}

 The units that we use are astronomical units (AU) for distance ($\sim 1.5$ $\times 10^{11}$ km), solar masses (M$_{\odot}$) for mass ($\sim$ 2 $\times 10^{30}$kg) and years (yr) for time ($\sim$ 3 $\times 10^7$). One astronmical unit is the distance between the Earth and the Sun and a solar mass is the mass of our Sun. In some cases we use time in terms of crossing time of a system where the crossing time is given by Equation \ref{eq:crtime} \citep{karttunen}. Here G is the gravitational constant, M is the total mass of the system and E$_0$ is the energy of the system.
\begin{equation}
T_{cr}=GM^{5/2}(2|E_0|)^{-3/2}
\label{eq:crtime} \end{equation}

ARCcode was provided by Prof. Seppo Mikkola \citep{Mikkola1993,Mikkola1996,Mikkola1999,Mikkola2002,Mikkola2006,Mikkola2008,Hellstrom2010,MikkolaElsev.2013,Mikkola2013} where we used up to 2.5$^{th}$ order pN corrections for 3-body simulations. The spin-orbit terms (which are presented at the 1.5 pN order \citep{Barker1975}) used in the numerical integration are explained in detail by \citet{Valtonen2010} and by \citet{Mikkola2020}. This is summarised briefly in \ref{code}. 

\subsection{Code Implementation}\label{code}

ARCcode conducts orbital numerical integration of the system of triplets based on initial conditions placed by the user. For this work we conduct simulations with post-Newtonian corrections up to the 2.5th order. 

This code performs orbital integration using a logarithmic Hamiltonian leapfrog method. The transformed equations of motion for both the time and co-ordinates are obtained from the logarithmic Hamiltonian (a full derivation is presented in \citet{Mikkola2013}). Using the leapfrog algorithmic method, these equations can be solved approximately.

The pN equations developed by \citet{MoraWill2004} and further explained in \citet{mikkolabook2020} which are used for integration (up to the 2.5$^{th}$ order) are listed here. Here, we consider two bodies \textit{m$_1$} and \textit{m$_2$} with separation, \textit{r} and relative velocity, \textit{v}, where $m = m_1 + m_2$; 
\begin{math}
    \eta =\frac{m_1m_2}{m^2}
\end{math} and \begin{math}
    \bf{n}=\frac{\bf{r}}{r}
\end{math}.

The pN equations used, up to 2.5$^{th}$ order, are as follows:

\begin{equation}
    A_1 = 2(2 + \eta)\left(\frac{m}{r}\right) - \left(1 + 3\eta\right)v^2 + 1.5\eta\dot r^2
\end{equation}

\begin{equation}
       A_2 = -\frac{3}{4}\left(12 + 29\eta\right)\left(\frac{m}{r}\right)^2 - \eta\left(3 - 4\eta\right)v^4 + \\\left(2 + 25\eta + 2\eta^2\right)\left(\frac{m}{r}\right)\dot r^2 + 1.5\eta\left(3 - 4\eta\right)v^2\dot r^2 
\end{equation}

\begin{equation}
    A_{2.5} = \frac{8}{5}\eta\frac{m}{r}\dot r\left(\frac{17}{3}\frac{m}{r} + 3v^2\right)
\end{equation}

\begin{equation}
 B_1 = 2\left(2 - \eta\right)\dot r
\end{equation}

\begin{equation}
    B_2 = -\frac{1}{2}\dot r[\left(4 + 41\eta + 8\eta^2\right)\left(\frac{m}{r}\right) \\ 
    -\eta\left(15 + 4\eta\right)v^2 + 3\eta\left(3 + 2\eta\right)\dot 
    r^2 ] 
\end{equation}

\begin{equation}
B_{2.5} = -\frac{8}{5}\eta\left(\frac{m}{r}\right)\left[3\left(\frac{m}{r}\right) + v^2\right]
\end{equation}

Then, the total effect on the acceleration of the bodies is given by the following:

\begin{equation}
   A_{tot} = A_1/c^2 + A_2/c^4 + A_{2.5}/c^5  
\end{equation}

\begin{equation}
    B_{tot} = B_1/c^2 + B_2/c^4 + B_{2.5}/c^5
\end{equation}

Which is then written in the acceleration term as:
\begin{equation}
    \bm{\ddot r} = \bm{\ddot r}_N - \frac{m}{r^2}\left(A_{tot}\bm{n} + B_{tot}\bm{v}\right) + \bm{\ddot x}_{SO} + \bm{x}_Q 
\end{equation}

The term, $\bm{\ddot x}_{SO}$ is the spin-orbit contribution and $\bm{\ddot x}_Q$ is the quadrupole-monopole term. Both describe the effect of spin on the acceleration of the body. From \citet{Barker1975} and described in \citet{Valtonen2010} these terms are defined as follows:

\begin{equation}
    \bm{\ddot x}_{SO} = \frac{Gm}{r^2}\left(\frac{Gm}{c^3r}\right)\left(\frac{1 + \sqrt{1 - 4\eta}}{4}\right)
    \times \chi \Bigg[\left(12\left[\bm{s}_1 \cdot (\bm{n}\times\bm{v})\right]\right)\bm{n}\\
    + \left[\left(9 + 3\sqrt{1 - 4\eta}\right)\dot r\right]\left(\bm{n}\times\bm{s}_1\right) 
    - \left[7 + \sqrt{1 - 4\eta}\right](\bm{v}\times\bm{s}_1)\Bigg]
\end{equation}

\begin{equation}
     \bm{\ddot x}_Q = \chi^2\frac{3G^3m^2_1m}{2c^4r^4}\left[\left(5(\bm{n}\cdot\bm{s_1})^2-1\right)\bm{n} - 2(\bm{n}\cdot\bm{s_1})\bm{s_1}\right]
\end{equation}

In the spin-orbit and quadrupole-monopole terms, $\chi$ represents the spin parameter and s$_1$ represents the spin of the black hole.

\subsubsection{Constraints}

ARCcode does not use cross terms which over relativistic timescales become necessary for describing the evolution of systems \citep{Will2014a,Will2014b}. However, for this study our aim is to see via the use of ARCcode, if there is any measurable effect of the inclusion of spin. Here, we focus on studying the influence of spin on the final outcomes of the triplets statistically. 

 We also limit the case to where only one of the black holes in our triplet system has spin. This is because ARCcode allows only one of the black holes to rotate within a given simulation. However, in numerical integration, if one black hole is much bigger than the others, then the spin of the smaller black hole is considered negligible \citep{mikkolabook2020}.  In our simulations, though the masses are on similar scales we always choose the largest black hole to be the rotating one of the three.

\subsection{Initial Conditions}

All black holes begin with zero initial velocity and only one of them is allowed to rotate - this is the biggest black hole in each triplet system. The magnitude of the spin vector is kept to $<$1 in normalised units as a spin magnitude $=$1 results in zero surface gravity of the black hole and the theoretical naked singularity which theoretically cannot exist \citep{Bardeen1973,Wald1999}. All bodies start with co-ordinates limited to the xy-plane. Timescales were kept as in \citet{Chitan2021}, with the crossing times of systems being utilized for comparisons of lifetimes.

\begin{figure*}[ht!]
    \centering
    \includegraphics[width=0.5\textwidth]{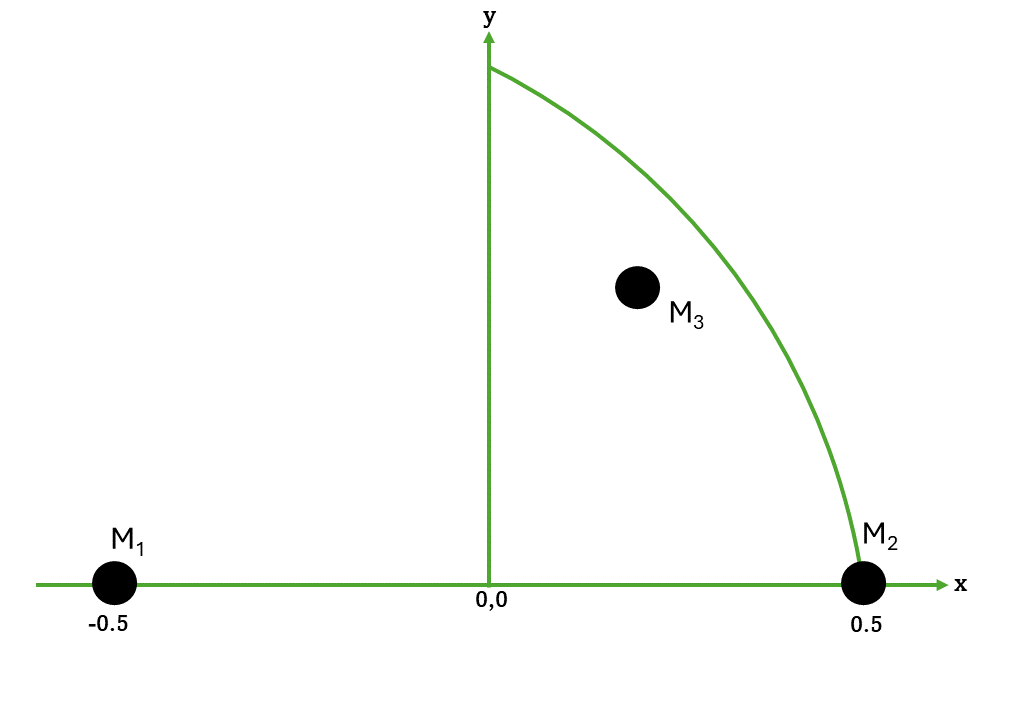}
    \caption{The Agekian-Anosova homology region (region D). This describes phase space for the zero initial velocity, equal mass three body system.}
    \label{fig:AAregion}
\end{figure*}

We focus first on the Agekian-Anosova homology region, also referred to as region D. This forms \textit{Set One} of our simulations. This region describes the phase space region for zero initial velocity, equal mass three body systems \citep{anosova1984,anosova1991,anosova1994}. In this space, two bodies are fixed at co-ordinates (0.5, 0) and (-0.5, 0) while the third body is placed in the region within the unit circle centered at (-0.5, 0) and bounded by the  x and y axes. This is shown in Figure \ref{fig:AAregion}. By varying the position of M$_3$ within this bounded region, a complete run of simulations is made. The bounded region was sampled by constructing 1000 circles centered at (-0.5,0) with decreasing radii. The upper limit for the radii was that of the initial unit circle and the lower limit was 0. Only quadrant one (x $>$ 0, y $>$ 0) was considered. Each of these circles were constructed of individual points with a stepsize of 0.01. These points for all of the circles created the set of positions in which the M$_3$ body was placed during simulations. For each run, there were 8094 different positions of M$_3$, making 8094 individual simulations per run.

In \textit{Set One}, we run simulations with all Schwarzschild black holes. Then, we re-run the simulations with a black hole M$_3$, having a spin magnitude near unity. To fully cover the geometrical  Agekian-Anosova region we also repeated simulations with M$_3$ fixed at (-0.5,0) and then fixed at (0.5,0). The results of these last two are provided in the Appendix. A summary of this is provided in Table \ref{tab:SetOne}.

\begin{table*}[ht!]
    \caption{A summary of Set One simulations. Following the setup shown in Figure \ref{fig:AAregion}. Each Run consists of 8094 individual simulations where M$_1$ and M$_2$ stay fixed on the negative and positive x-axis, respectively, but M$_3$ moves each time within the bounded region. An initial run was conducted where none of the black holes had spin, as a control. To cover the Agekian-Anosova region fully, we also conduct runs three and four where the spinning black hole is fixed at M$_1$ and M$_2$, respectively.  Black hole mass is fixed at 10$^{6}$ M$_{\odot}$, with the spinning black hole being more massive at 10$^{7}$ M$_{\odot}$ for each case.}
    \begin{center}
        \begin{tabular}{||c|c|c|c|c||}
        \hline
         Run & Black Hole Position   &  Spin Magnitude & Mass (10$^{6}$ M$_{\odot}$) & (x,y) Position\\
         \hline
         \multirow{3}{4em}{One} & M$_1$ & 0  & 1 & (-0.5,0) \\
         
        & M$_2$    & 0 & 1 & (0.5,0) \\
         
         & M$_3$     & 0 & 10 & varying within region D \\
         \hline
         \multirow{3}{4em}{Two} & M$_1$ & 0 & 1 & (-0.5,0)\\
         & M$_2$ & 0 & 1 & (0.5,0)\\
         & M$_3$ & 0.95 & 10 & varying within region D \\
         \hline
         \multirow{3}{4em}{Three} & M$_1$ & 0.95 & 10 & (-0.5,0)\\
         & M$_2$ & 0 & 1 & (0.5,0)\\
         & M$_3$ & 0 & 1 & varying within region D\\
         \hline
         \multirow{3}{4em}{Four} & M$_1$ & 0 & 1 & (-0.5,0)\\
         & M$_2$ & 0.95 & 10 & (0.5,0)\\
         & M$_3$ & 0 & 1 & varying within region D\\
         \hline
        \end{tabular}
    \end{center}
    \label{tab:SetOne}
\end{table*}

\textit{Set Two} of our simulations come from the Pythagorean triangles and Burrau's problem. We conducted simulations in depth simulations with the classical 3,4,5 triangle including both increasing spin and increasing mass. The typical 3,4,5 triangle is provided in Figure \ref{fig:3,4,5}. The black holes are initialised such that their masses are multiples of the side of the triangle they are opposite from. When we configure the triangles, the masses of the black holes that we are studying are just multiplied by the value given by the particular triangle. The most massive black hole is the one that spin is added to. The distances, shown as `d' in Figure \ref{fig:3,4,5}, is kept at 206265 AU in our simulations. This value is equivalent to 1 parsec.
\begin{figure*}[ht!]
    \centering
    \includegraphics[width=0.5\textwidth]{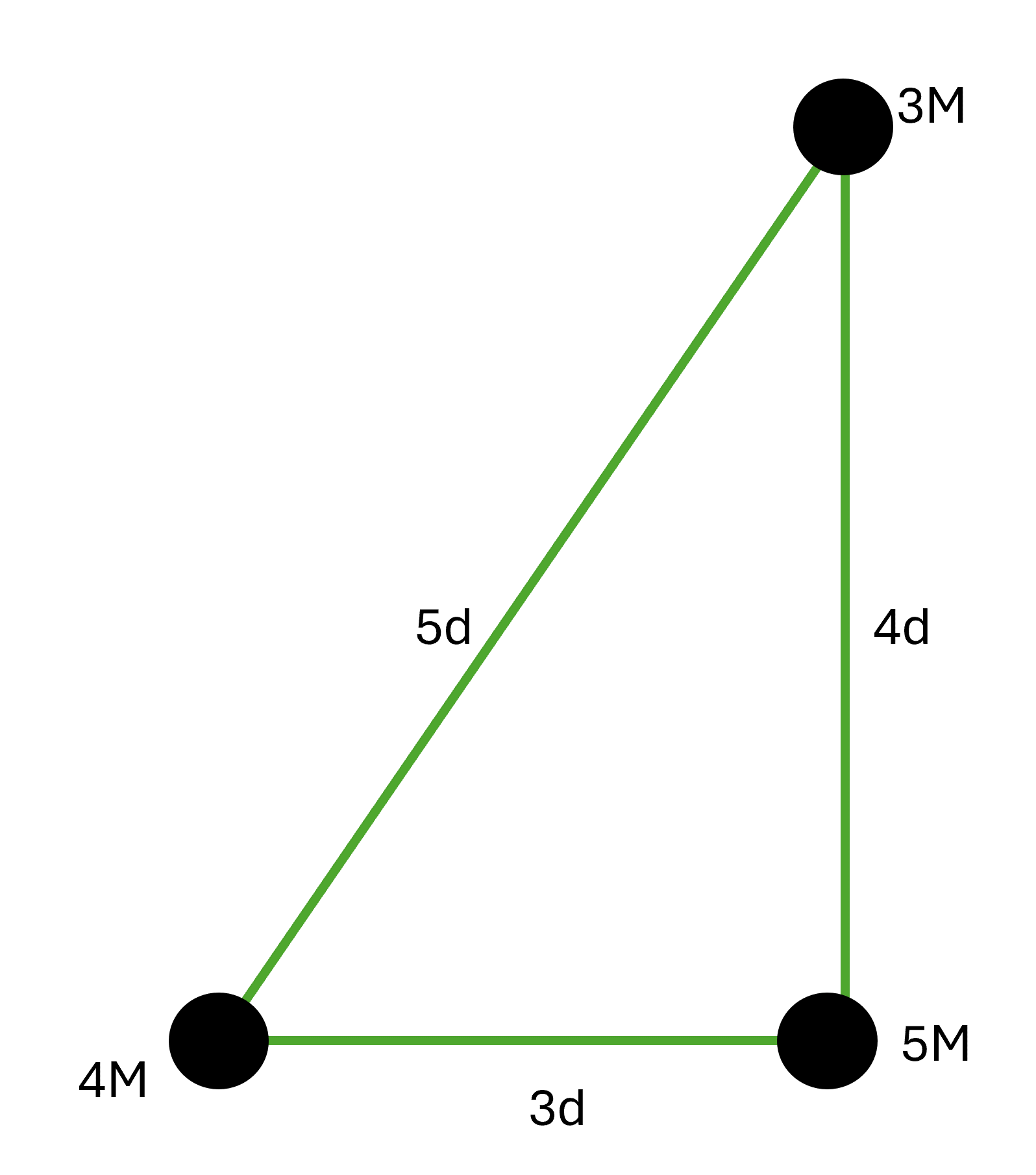}
    \caption{The Pythagorean 3,4,5 triangle showing the initialised positions of black holes.}
    \label{fig:3,4,5}
\end{figure*}

In our previous study \citep{Chitan2021} we used an informal method of classifying the triangles into either more hierarchical or less hierarchical depending on the angle between the hypotenuse and the second longest side of the triangle. If there was an angle $>$ 30$^{\circ}$, the triangle was considered a non-hierarchical triangle and if the angle was $<$ 30$^{\circ}$, a hierarchical triangle. This is shown in Table \ref{tab:angles}.
\begin{table}[ht!]
\caption{Pythagorean triangles classified into hierarchical or non-hierarchical.}
\begin{center}

    \begin{tabular}{|c|c|c|}
    \hline
    Triangle & Classification  & Angle ($^{\circ}$)  \\
    \hline
         3,4,5 & Non-Hierarchical & 36.87\\
      \hline
      5,12,13   &Hierarchical  & 22.62\\
      \hline
       7,24,25& Hierarchical  & 16.26 \\
      \hline
       8,15,17& Hierarchical  & 28.07 \\
      \hline
       9,40,41& Hierarchical  & 12.68\\
      \hline
       11,60,61& Hierarchical  & 10.39\\
      \hline
       12,35,37& Hierarchical  & 18.93 \\
      \hline
       13,84,85& Hierarchical  & 8.80 \\
      \hline
       16,63,65& Hierarchical  & 14.25 \\
      \hline
          20,21,29& Non-Hierarchical &  43.60\\
          \hline
         28,45,53& Non-Hierarchical & 31.89\\
          \hline
          33,56,65& Non-Hierarchical & 30.51\\
          \hline
        36,77,85& Hierarchical  & 28.07 \\
      \hline
        39,80,89& Hierarchical  & 28.07 \\
      \hline
         48,55,73 & Non-Hierarchical& 41.11\\
          \hline
         65,72,97 & Non-Hierarchical& 42.08\\
         \hline
    \end{tabular}
    \end{center}
\label{tab:angles}
\end{table}

The effect of increasing spin was the focus for Set Two. We were interested in seeing how the effect of spin varied across the different masses of the black holes. To do this we constructed three different runs. In the first of run of set two simulations, we increased the mass of the black holes from 10$^{0}$ M$_{\odot}$ to 10$^{12}$ M$_{\odot}$ in increasing factors of 10. For the second run we zoomed in more closely to the mass range of 10$^{4}$ M$_{\odot}$ to 10$^{7}$ M$_{\odot}$ in increasing factors of 10$^{0.2}$ M$_{\odot}$. For the third run we focused only on the 10$^{5.5}$ M$_{\odot}$ mass since this mass was found to be the mass of our black hole triplets at which triplets preferred a merger outcome to an escape outcome \citep{Chitan2021}. For all of these runs, the most massive of the black holes, based on the Pythagorean triangles they were set up with. This spin varied from [0,0,0] to [0.55,0.55,0.55] increasing with [0.05,0.05,0.05]. These were kept along the diagonal for simplicity and not to introduce extra variability in an already complex system. The summary of the Set Two simulations is given in Table \ref{tab:SetTwo}.

\begin{table*}[ht!]
\caption{A summary of Set Two simulations. This is from the Pythagorean triangles. For each of the Runs listed, all Pythagorean triangles with a hypotenuse $<$ 100 pc are used while the masses of the black holes are increased as listed.}
    \begin{center}
        \begin{tabular}{||c|p{0.2\linewidth}|p{0.4\linewidth}||}
        \hline
         Run & log$_{10}$(Mass (M$_{\odot}$))   &  Spin Vector\\
         \hline
         One & 0, 1, 2, 3, 4, 5, 6, 7, 8, 9, 10, 11 & [0 0 0], [0.05 0.05 0.05], [0.1, 0.1, 0.1], [0.15, 0.15, 0.15], [0.2, 0.2, 0.2], [0.25, 0.25, 0.25], [0.3, 0.3, 0.3], [0.35, 0.35, 0.35], [0.4, 0.4, 0.4], [0.45, 0.45, 0.45] [0.5, 0.5, 0.5], [0.55 0.55 0.55]  \\ 
         \hline
         Two & 4, 4.2, 4.4, 4.6, 4.8, 5, 5.2, 5.4, 5.6, 5.8, 6, 6.2, 6.4, 6.6, 6.8, 7 &  As above\\     
         \hline
        Three & 5.5 &  As above\\
        \hline
         
        \end{tabular}
    \end{center}
    \label{tab:SetTwo}
\end{table*}

\subsection{Parameters}

The main parameters we consider are: lifetimes of the systems, fraction of mergers, evolution of some of the orbital elements and how the motion is transformed from two-dimensional to three-dimensional.  
Lifetime defines how long a system takes to evolve from three black holes into two. This can happen if two of the black holes merge (merger) or one of the black holes escapes the center of mass of the triplet (escape). The systems are always initialised in the 2D plane, but as we include spin this takes the system to 3D. We then follow the evolution of the system along the z-axis. For the orbital elements, we study this in Set Two. These are the orbital elements of one of the black holes with respect to the most massive, rotating black hole.

\section{Results and Discussion}\label{sect: Section 3}

\subsection{The Effect of Spin in Set One}

The initial conditions were set so that two black holes of 1$\times$10$^6$ M$_{\odot}$ each were fixed on the x-axis while the position of the third black hole of 10$\times$10$^6$ M$_{\odot}$ was varied within the bounded region.This is described in Table \ref{tab:SetOne} where 8094 individual simulations were conducted for each scenario: with spin and without spin.

\begin{figure}[ht!]
    \centering
    \includegraphics[width=0.75\textwidth]{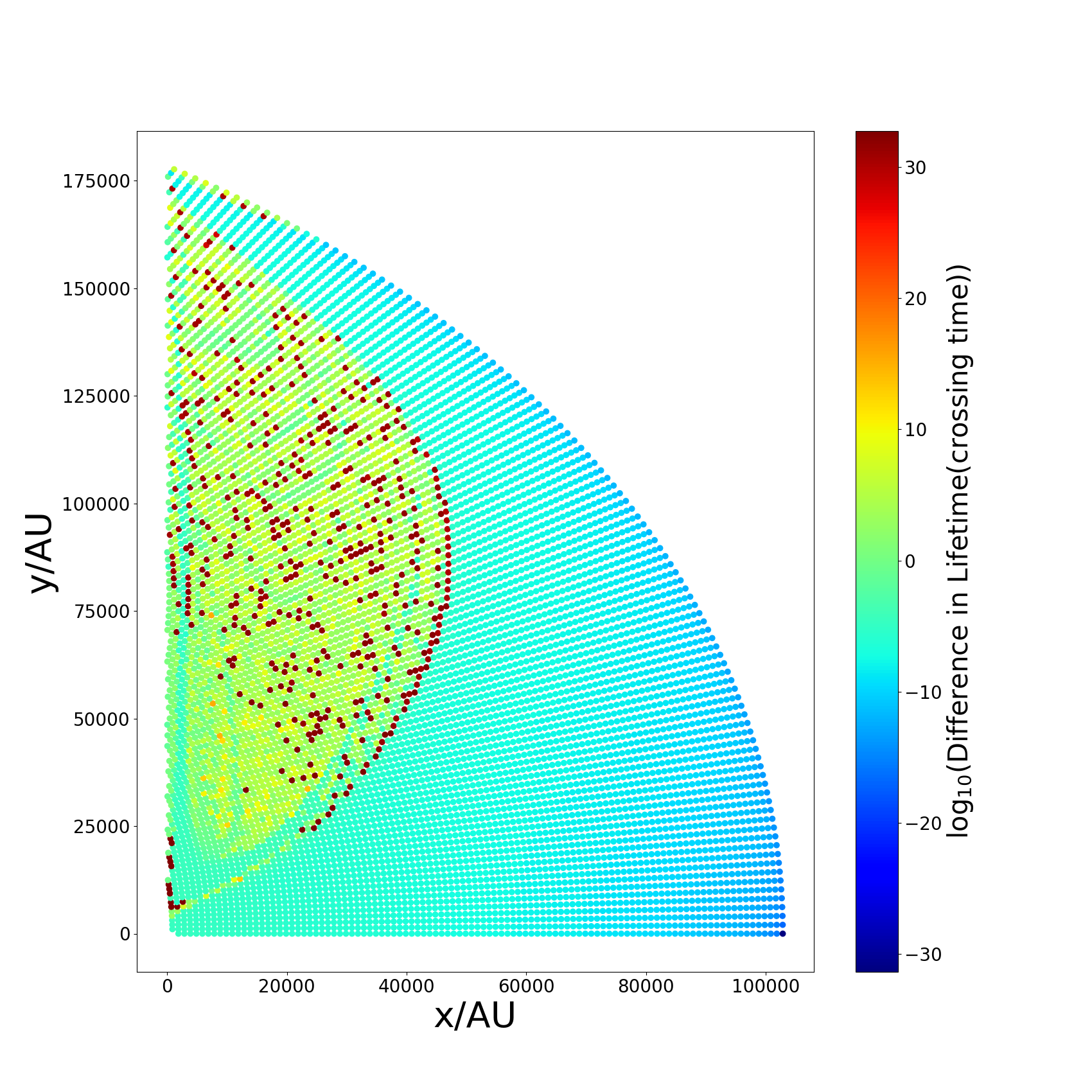}
    \caption{The difference in log$_{10}$(lifetime/crossing time) between simulations in the Agekian Anosova region. This is the difference between Run One and Run Two from Table \ref{tab:SetOne}. Here the masses of the fixed black holes are 10$^6$ M$_{\odot}$. The spinning black hole varies within the region D in every simulation and has a mass of 10$\times$10$^6$ M$_{\odot}$.}
    \label{fig:lifespindiff}
\end{figure}

 Figure \ref{fig:lifespindiff} shows the difference in lifetimes between the cases with the inclusion of spin and without it. For this figure, the x and y axis show the initialised position of the black hole with spin, while the colour indicates the length of the lifetime for that particular setup. Here, a clearly defined region (the green portion) is seen. The addition of spin in the simulations can result in lifetimes of triplet systems increasing in some cases 30 magnitudes of order. This region of high variability corresponds to more non-hierarchical triplets. In these non-hierarchical triplets, there are initial configurations that are very similar but live magnitudes longer. This corresponds to the red dots within the green region. The chaoticity of the three-body problem is most likely the cause of this. Very small changes in the initial conditions result in very large difference in outcomes. In some cases here, almost identical setups have a lifetime difference of 10$^{20}$ years. The effect of the additional element of rotation on M$_3$ is likely due to the high sensitivity of these non-hierarchical triplets. Similar results are found when we analyse number of mergers and number of two-body encounters.

\begin{figure}[ht!]
    \centering
    \includegraphics[width=0.8\textwidth]{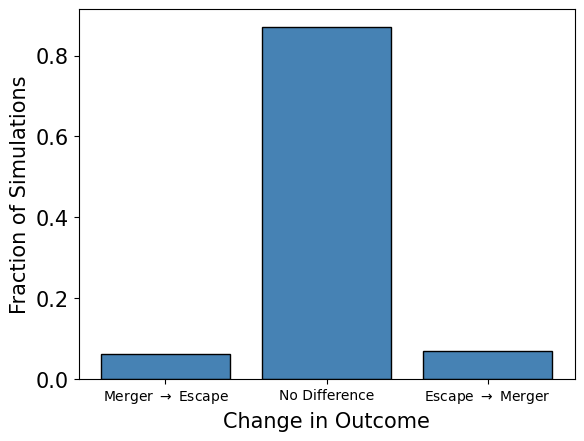}
    \caption{The change of the triplet system after spin is added into Set One simulations. Merger means the system ends when two black holes have merged, escape means that one of the black holes escapes the triplet center of mass.}
    \label{fig:outcomechange}
\end{figure}

 The outcomes of the black hole triplet system is also observed to change with the addition of spin in 12.9\% of Set One simulations. The fraction of simulations that end via, merger, via escape, or do no change with the addition of spin, is shown in Figure \ref{fig:outcomechange}.

\subsection{The Effect of Increasing Spin in Set Two}
\begin{figure}[ht!]
    \centering
    \includegraphics[width=0.8\textwidth]{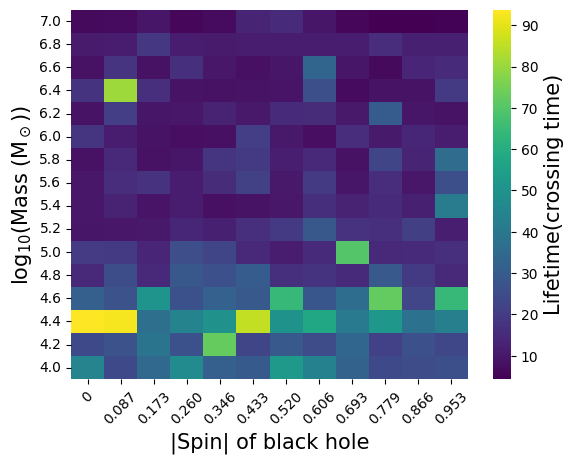}
    \caption{The average lifetime, in crossing time, for all Pythagorean triangular configurations within the mass range of 10$^{4}$ M$_{\odot}$ - 10$^{7}$ M$_{\odot}$ and with increasing spin magnitude.}    \label{fig:avglife4-7}
\end{figure}

Here we focus on the ensemble of Pythagorean triangular configurations (Set Two). When we consider the average lifetime of the triplet systems across all mass ranges ($10^0$ M$_\odot$ - $10^{12}$ M$_\odot$) in Set Two Run One (Refer to Table \ref{tab:SetTwo}), there is no discernible effect of spin above the mass 10$^8$ M$_\odot$. Mass becomes the dominant factor with large mass systems living for shorter times, which is expected. However, the intermediary mass range we observe variation in the lifetimes. In Figure \ref{fig:avglife4-7}, this mass range is looked at more closely from Set Two Run Two. It may be that spin has a small effect on the lifetimes around the 10$^{4}$ M$_{\odot}$ - 10$^{5}$ M$_{\odot}$ mass range.

\begin{figure}[ht!]
    \centering
    \includegraphics[width=0.8\textwidth]{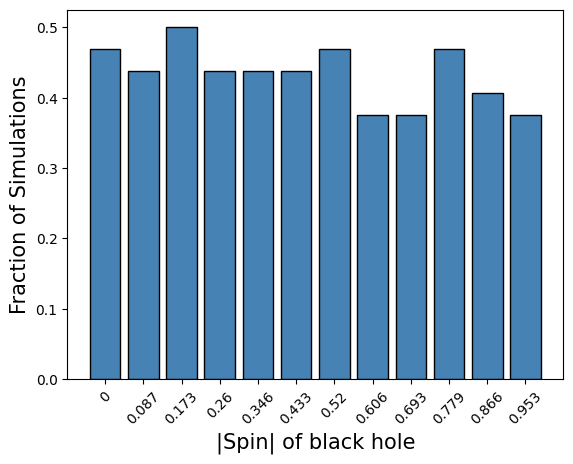}
    \caption{The fraction of mergers that take place among all the triangles at 10$^{5.5}$ M$_{\odot}$ with increasing spin magnitude of the most massive black hole in each system.}
    \label{fig:fractmerge}
\end{figure}

For the mass of 10$^{5.5}$ M$_{\odot}$ (Set Two Run Three), we look at the fraction of mergers occurring at each increasing spin value. In \citet{Chitan2021}, we found that this mass was important within the Pythagorean triplet dataset because at this mass there was a transition from escape dominated dynamics (triplet lifetimes ending when one black hole escaped) to merger dominated dynamics (triplet lifetime ending when two black holes merged). Because of this, we pay special attention to this mass to see if spin influences a change. For each of the 16 triangular configurations, two mergers are possible allowing 32 total merger events possible per spin value of $M_1$. The results are shown in Figure \ref{fig:fractmerge}. The fraction of mergers stay around the 40$\%$ mark. From our previous study, at 10$^{5.5}$ M$_{\odot}$, the fraction of mergers was at 50$\%$. While the difference is very subtle, it may be that adding spin helps to facilitate escape over merger. From this it appears that spin does not have too strong of an influence on the outcomes of the Set Two simulations. 

\begin{figure}[ht!]
    \centering
    \includegraphics[width=0.8\textwidth]{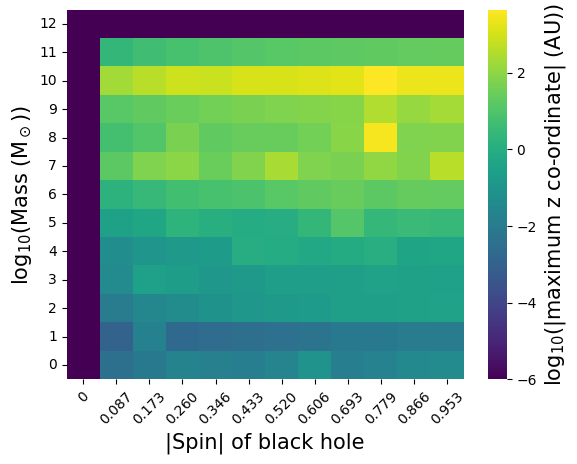}
    \caption{The log$_{10}$(average $|$maximum z$|$ co-ordinate/AU) of the rotating black hole throughout the lifetime of the triplet system as both mass and spin are varied for all triangular configurations.}
    \label{fig:allmaxz}
\end{figure}

One influence of spin is that it naturally transforms the 2D motion of the system into 3D. Initially, all bodies begin with co-ordinates limited to the xy-plane, however the spin vector of the rotating black hole is in the form [x y z]. We first consider the average among all triangles. There is an observable shift from the xy-plane to the xyz-plane upon the addition of spin illustrated in Figure \ref{fig:allmaxz}, which shows the average maximum z co-ordinate achieved by the rotating black hole, while the triplet system is intact (i.e. before any merger or escape). This z-motion is normalised to the hypotenuse for each triangle. The added spin has the greatest effect on the motion in the z-axis at the large masses, while intermediary masses see some effect as well. This is interesting as the previously predictable large mass cases (such as 10$^{10}$ M$_{\odot}$, 10$^{11}$ M$_{\odot}$), obtain an added layer of complexity. However, the final outcomes such as lifetimes, mergers and the number of two-body encounters are not really affected by this as previously discussed.

 At the smaller mass units (10$^{0}$ M$_{\odot}$-10$^{2}$ M$_{\odot}$), the influence of spin is not very apparent but as the systems increase in mass, there is an increase in the maximum z-coordinate achieved by the rotating black hole. This falls within the mass range of $~$ 10$^{4}$ M$_{\odot}$ - 10$^{10}$ M$_{\odot}$. As the mass unit increases beyond this, there is still 3-dimensional motion but much less than the intermediary masses. 

\begin{figure}[ht!]
     \centering
     \begin{subfigure}[b]{0.45\textwidth}
         \centering
         \includegraphics[width=\textwidth]{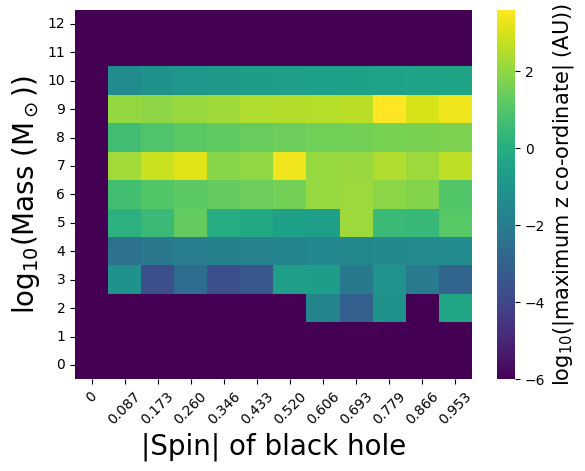}
         \caption{3,4,5 triangle}
         \label{fig:345max}
     \end{subfigure}
     \begin{subfigure}[b]{0.45\textwidth}
         \centering
         \includegraphics[width=\textwidth]{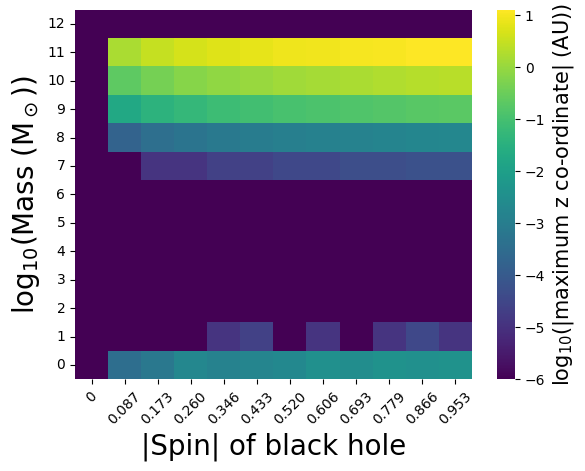}
         \caption{13,84,85 triangle}
         \label{fig:138485zmotion}
     \end{subfigure}
        \caption{The lg($|$maximum z$|$ co-ordinate/AU) of the spinning black hole throughout the lifetime of the triplet system as both mass and spin are varied for the 3,4,5 triangular configuration (a) and the 13,84,85 triangular configuration (b).}
        \label{fig:maxZcompare}
\end{figure}

For Burrau's 3,4,5 triangle which is considered chaotic - there is an observable effect of spin at intermediary and large mass units as seen in Figure \ref{fig:maxZcompare} (a). Smaller masses are not as influenced to move in the z-axis and larger masses end quickly before this type of motion can present itself. 

Such behaviour is observed in the \NHT but more minimally in the \HT at the smaller mass units. Interestingly, at larger mass units, the \HT exhibit some z-motion before quick mergers take place as can be seen in Figure \ref{fig:maxZcompare} (b), where the results of the 13,84,85 triangle are shown. It is also more noticeable here that as the spin magnitude increases, the maximum z-coordinate also increases. The major difference is evident in how much the body the rotating black hole moves along the z-axis. For \NHT, the z value can become much larger than for the \HT. 

These differences are typical as non-hierarchical configurations are thought to be more chaotic and unpredictable than the hierarchical ones \citep{Anosova1990,Anosova1992}. In all triangles at the extreme mass unit of 10$^{12}$ M$_{\odot}$, systems end almost immediately without enough time to develop motion in the z-axis. These are the bands that appear in Figures \ref{fig:allmaxz} and \ref{fig:maxZcompare}. 

\begin{figure*}
    \centering
    \includegraphics[width=0.5\textwidth]{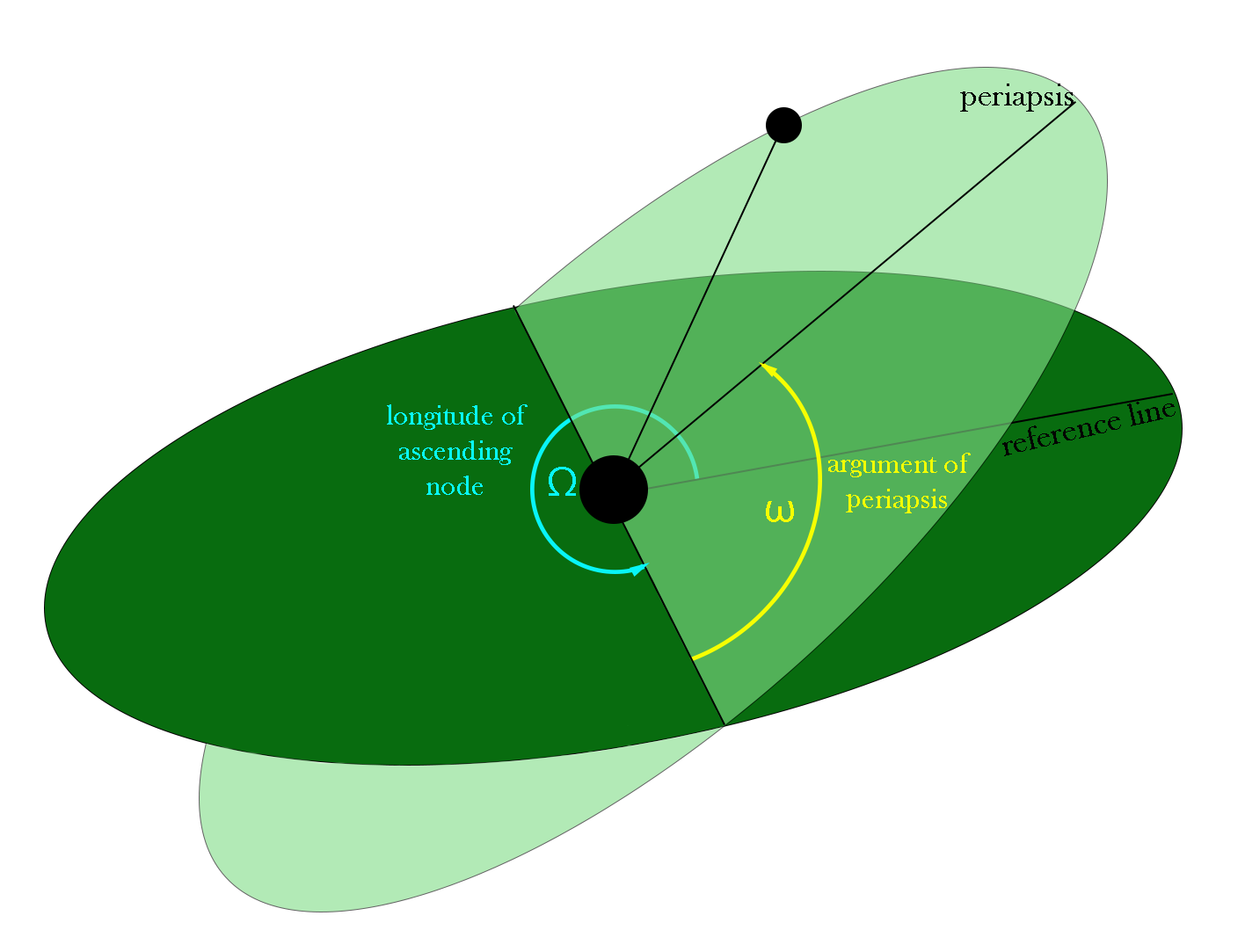}
    \caption{The argument of periapsis, $\omega$ and the longitude of ascending node, $\Omega$ shown on an orbital diagram.}
    \label{fig:orbital}
\end{figure*}

From the Set One simulations, we also observe an effect on the orbital elements of our systems. The argument of periapsis, $\omega$, and the longitude of ascending node, $\Omega$ are affected by the added spin in the system. These orbital elements are shown in Figure \ref{fig:orbital}. Periapsis is the point of closest approach in the orbit. The angle between the line of periapsis and where the orbital plane intersects the reference plane (light green and dark green respectively in Figure \ref{fig:orbital}), is the argument of periapsis. This is shown in yellow in Figure \ref{fig:orbital}. The longitude of the ascending node is the angle between the origin of the longitude of the reference plane (black) and the intersection of the orbital plane with the reference plane as the orbiting body ascends. This is shown in blue in Figure \ref{fig:orbital}. We have observed that in some cases of this study, these two parameters fluctuate.

\begin{figure}
     \centering
     \begin{subfigure}[b]{0.45\textwidth}
         \centering
         \includegraphics[width=\textwidth]{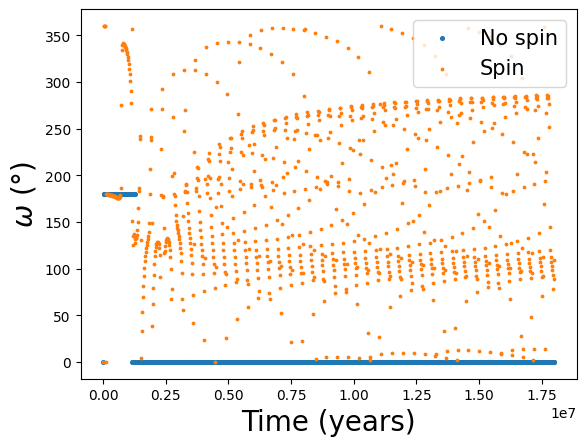}
         \caption{20,21,29 triangle}

     \end{subfigure}
     \begin{subfigure}[b]{0.45\textwidth}
         \centering
         \includegraphics[width=\textwidth]{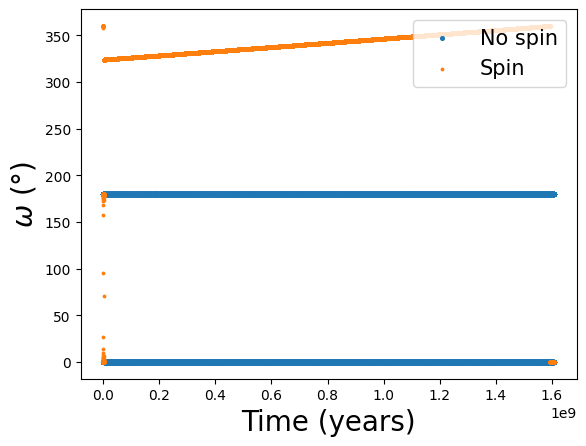}
         \caption{9,40,41 triangle}

     \end{subfigure}
     \begin{subfigure}[b]{0.45\textwidth}
         \centering
         \includegraphics[width=\textwidth]{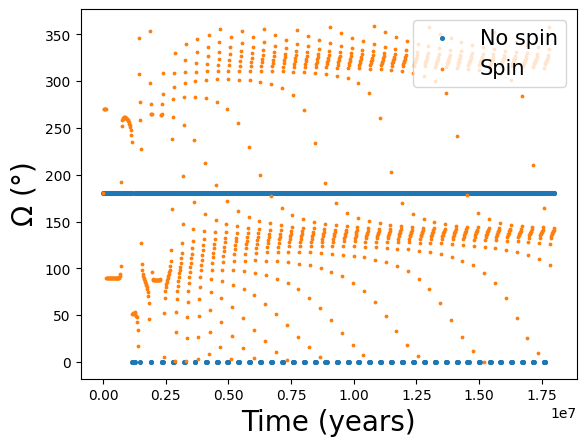}
         \caption{20,21,29 triangle}

     \end{subfigure}
     \begin{subfigure}[b]{0.45\textwidth}
         \centering
         \includegraphics[width=\textwidth]{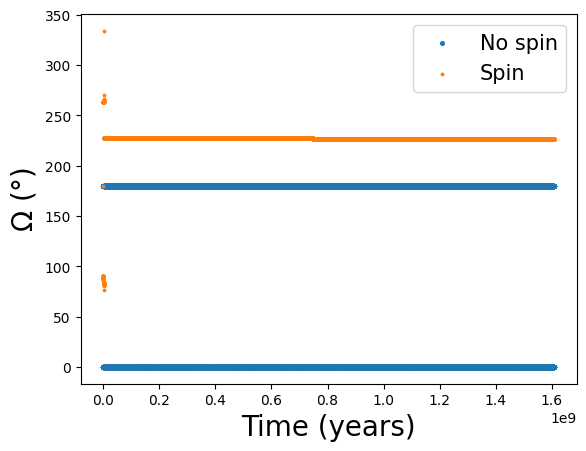}
         \caption{9,40,41 triangle}

     \end{subfigure}
        \caption{The change in $\omega$, the argument of periapsis, (top) and $\Omega$, the longitude of ascending node, (bottom) for the 20,21,29 (left) configuration vs. the 9,40,41 configuration (right). In all cases the masses of the black holes are maintained at 10$^{4.5}$ M$_{\odot}$ within each triangular configuration.}
        \label{fig:orbitalevolution}
\end{figure}

As with the motion in the z-axis, these variations are mainly seen with the \NHT. In Figure \ref{fig:orbitalevolution}, the panels on the left show the variation in $\omega$ and $\Omega$ for the 20,21,29 configuration and those on the right for 9,40,41 configuration. These represent the orbital elements of black hole M$_3$ (non-rotating) with respect to M$_1$ (rotating) for intermediary the mass unit of 10$^{4.5}$ M$_{\odot}$, the mass at which we see the most influence of spin in the lifetime of our systems.  In blue are the orbital elements over time when there is no spin in the system. The orange represents the orbital elements when M$_1$ has maximal rotation of [0.55 0.55 0.55]. Here, it can be seen that the orbital elements express much more variation over time for the \NHT  than the \HT  shown. This motion may be more complex for the \NHT due to longer lifetimes and more chaotic motion. For the simulations without spin, there is still slight variation in periapsis which we expect from relativistic precession of the periapsis.

\begin{figure*}
    \centering
    \includegraphics[width=\textwidth]{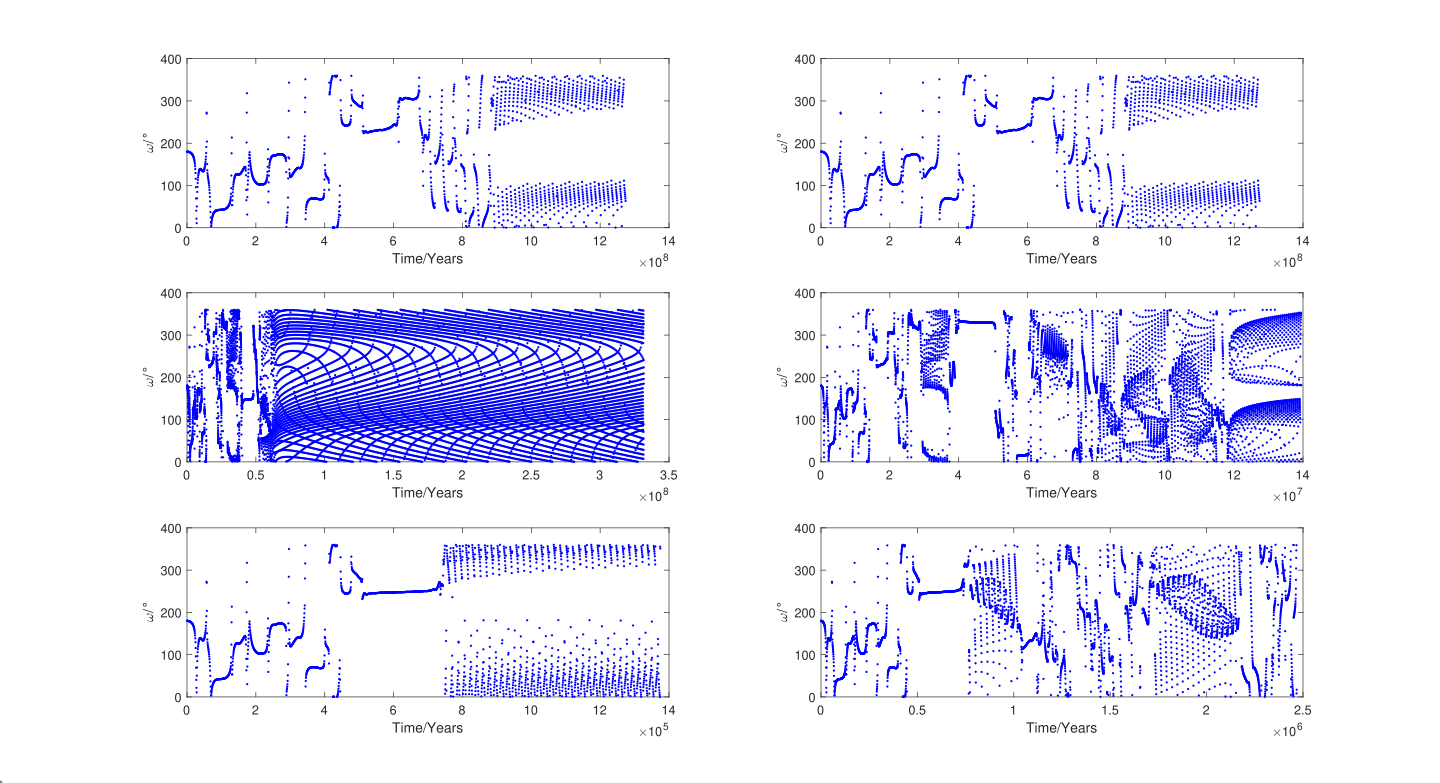}
    \caption{The variation in $\omega$ as both mass unit of the system and spin magnitude of M$_1$ (rotating black hole) are varied for the classical 3,4,5 configuration. On the left, the spin vector of M$_1$  is [0.1 0.1 0.1] and on the right, [0.55 0.55 0.55]. The top, middle and bottom rows show, respectively, 10$^{0}$ M$_{\odot}$, 10$^{3}$ M$_{\odot}$ and 10$^{6}$ M$_{\odot}$ mass units. }
    \label{fig:345massandspin}
\end{figure*}

From Figure \ref{fig:345massandspin}, we consider the motion of the argument of periapsis between the bodies M$_1$ (the rotating black hole) and M$_3$ (non-rotating). This is for Burrau's classical 3,4,5 configuration. The most interesting case is that of 10$^{3}$ M$_{\odot}$, with spin = [0.1 0.1 0.1]. Both the spin magnitude and the mass of black holes affect the argument of periapsis. Each of these systems, barring the 10$^{6}$ M$_{\odot}$ mass unit at [0.55 0.55 0.55] spin, ends in the escape of body M$_3$. At the 10$^{0}$ M$_{\odot}$, it appears as if there is no visible difference in $\omega$ between each. The differences appear at the larger mass units of 10$^{3}$ M$_{\odot}$ and 10$^{6}$ M$_{\odot}$. 
Of course, the motions of these orbital elements occur on very large timescales and it may not be measurable in the real setting.

For \HT, this precession is only noticeable at small mass units. For \NHT, however, this appears at small, intermediate and large mass units.

\section{Discussion and Conclusion}

From this study we have looked at the effect of spin in black hole triplets by conducting two sets of simulations. In Set One we looked at the Agekian-Anosova homology region (or region D) and in Set Two we looked at an ensemble of Pythagorean triangles. The results of Set One show that $\sim$12.9\% of simulations change how they end their lifetimes (from merger to escape or vice versa) with the inclusion of spin in this region. While we have touched on a few limitations of this study (outlined in Section \ref{sect:2}), this change in outcome of systems due to the added influence of spin is one that can give us an idea of how to treat real systems. In $>$ 87$\%$ cases, while the short term evolution might be unpredictable, when we add spin, the final, most dramatic description of the system -- whether the triplet breaks up due to a merger or an escape -- does not change. We must use caution however, as the intermediary behaviour of these systems do indeed change, in some cases quite significantly. This was seen when we looked at the differences in lifetimes of our systems and neighbouring points had orders of magnitude difference in system lifetimes (Figure \ref{fig:lifespindiff}). When we observe these real life triplets this sensitivity should be considered. This is  especially so if we try to predict mergers or escapes of supermassive black holes that are a part of triplets.

In Set Two, we have found that while there is not a significant effect of spin on some parameters like the fraction of mergers, the lifetimes of these systems may have been affected, particularly for intermediate mass black holes in the range 10$^{4}$ M$_{\odot}$-10$^{5}$ M$_{\odot}$.

We have also found that certain effects are more visible in the non-hierarchical triangles than in the hierarchical triangles like the motion of the orbital elements $\omega$, the argument of periapsis, and $\Omega$, the longitude of ascending node, owing to the chaotic tendencies of the non-hierarchical configurations. The inclusion of spin which is not perpendicular to the orbital plane leads to 3D motion.  This is of course, constrained to the distance unit of 1 parsec (206265 AU), which in larger mass cases, becomes more theoretical. 

\citet{karttunen} showed that when initial velocities were kept to zero, the motion of triplets was restricted to the plane. They derived statistical distributions from numerical orbital calculations of the binary energy within a triplet system for both the three-dimensional case and the planar case. They found that for each case the distributions were different statistically which could then affect, close encounters and lifetimes. Here, from our results (Figures \ref{fig:allmaxz}), this is corroborated as the inclusion of spin of one of the black holes results in three dimensional motion. This of course, makes the systems more complex and their behaviour even more chaotic and unpredictable. 

In the real astrophysical context, the effect of spin in binary systems have been studied extensively. \citet{Stella2009} looked at compact object binary systems and the astrophysical environments around these compact systems. The surrounding accretion disks of black holes and neutron stars should be affected by the strong gravity and the LTE of these rotating bodies. These effects should be large and noticeable when looking at X-ray flares from the hot accretion disks. The LTE then could potentially be used to determine certain characteristics about host compact objects. \citet{Valtonen2010} have also studied in great detail the OJ287 system. They concluded that the spin of the primary SMBH of the OJ287 system must influence the orbit of the secondary black hole around it. Using this, they were able to estimate what the spin of the primary SMBH should be. They later observed the flares from this system and were able to conclude that the spin is 0.313 \citep{Valtonen2016}. 

It is with no question that these binary systems are far less complicated than a triplet system. Tracing the orbits and predicting the spin based on frame-dragging effects are difficult, but feasible for binaries. For the triplet system, these effects, especially on the more \NHT appear to be greatly chaotic and unpredictable. The overall effect of the rotating black hole appears negligible when considering the final states of these systems i.e. mergers and two-body encounters. But, utilizing the orbits of these bodies and the potentially observable LTE on the other two bodies of these types of systems to glean information about the spin magnitude of a black hole may be quite difficult. The system may not be constrained to the orbital plane and orbital elements oscillate largely, all of which contribute to unpredictable outcomes.

For triplets of SMBH, which are expected to form hierarchically via galactic mergers \citep{Valtonen1996}, this type of orbital element oscillation may not appear, which can limit unpredictability. This coincides with simulations as \HT merge much more quickly, at smaller mass units, than the \NHT and therefore behave less chaotically. SBH triplets in the hierarchical formation however may experience this type of motion. Black holes of masses 1 M$_{\odot}$- 3 M$_{\odot}$ are considered to be highly theoretical. There have been a few candidates for the smallest black hole discovered. One of these was found by \citet{lam2022} via gravitational lensing, where the object could either be the smallest black hole found yet or a neutron star of mass 1.6 $-$ 4.4 M$_{\odot}$. When \citet{sahu2022} observed this object however they estimated a mass of around 7 M$_{\odot}$. \citet{Lam2023} reanalysed this object and found that it is closer to 6 M$_{\odot}$. \citet{Raidal2019} also studied primordial black holes in the mass range of 0.1 M$_{\odot}$-10$^{3}$ M$_{\odot}$ and their merger rates for potential LIGO detections. In some cases they found that a single primordial black hole nearby the binary would be the main disruption of a merger event. Here we find that for the most part, small mass triplet systems end in escape over merger with a remnant binary pair.

Again, differences in evolution of the system due to mass is observed. The effect of spin is strongest when looking at mass units of the intermediary ranges because it is at these mass units that lifetimes are longer than the large mass systems and black holes are large enough for their to be substantive triplet interaction. They interact more strongly for long time periods such that the effect of added spin is more evident. Of course, this is also restricted to the distance unit, fixed at 1 parsec (206265 AU). The influence of masses and initial distances have already been discussed in more detail in \citet{Chitan2021}.


\section*{Data Statement}
The data underlying this paper is available at 
\url{https://doi.org/10.6084/m9.figshare.19735960.v2} 

Code by Prof. Mikkola is available at - \url{http://www.astro.utu.fi/mikkola/}

Additional code by authors can be found here - \url{https://doi.org/10.6084/m9.figshare.13194146.v1}

\appendix
\section{Agekian-Anosova Region Extended}
\label{app1}
To fully cover all the possible geometrical configurations, we also conduct simulations with spinning black hole first, initialised at (-0.5, 0), Run Three, and then initialised at (0.5, 0), Run Four. In both cases, the masses of all the bodies remain as previously described and the spin magnitude on the spinning black hole also is kept at the same value. 

\begin{figure}[ht!]
    \centering
    \includegraphics[width=0.5\textwidth]{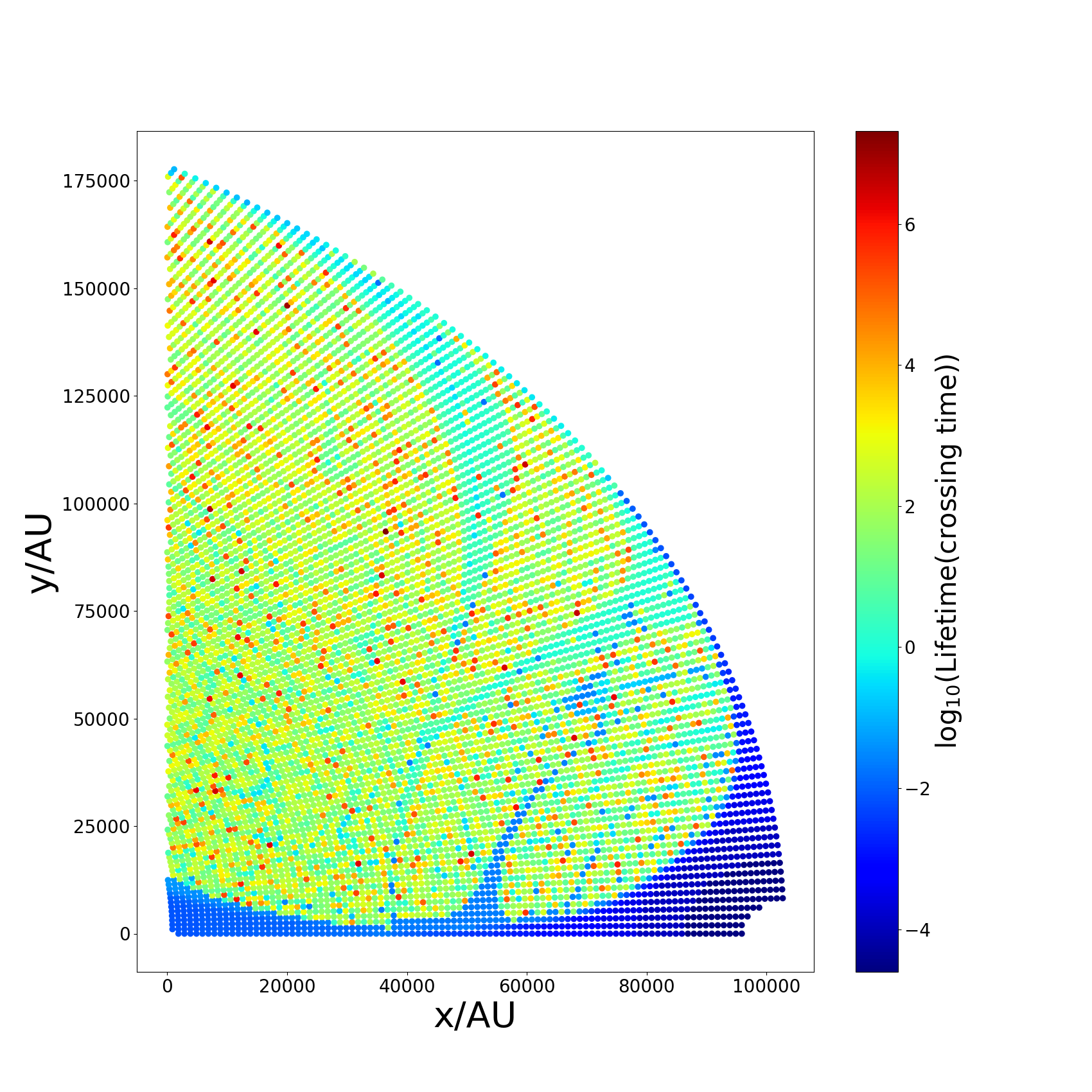}
    \caption{The log$_{10}$(lifetime/crossing time) for simulations in the Agekian Anosova region for Run Three of Set One simulations. See Table \ref{tab:SetOne}.}
    \label{fig:lifenegativeaxis}
\end{figure}

For the case where the spinning black hole is kept fixed on the negative axis, the log$_{10}$(lifetime/crossing time) looks like Figure \ref{fig:lifenegativeaxis}. The results in this case appear much different than in the other cases as most of the simulations are longer living. This is most likely due to the positioning of the most massive, rotating hole further away from the other two black holes as compared to the previous cases.  

\begin{figure}[ht!]
    \centering
    \includegraphics[width=0.5\textwidth]{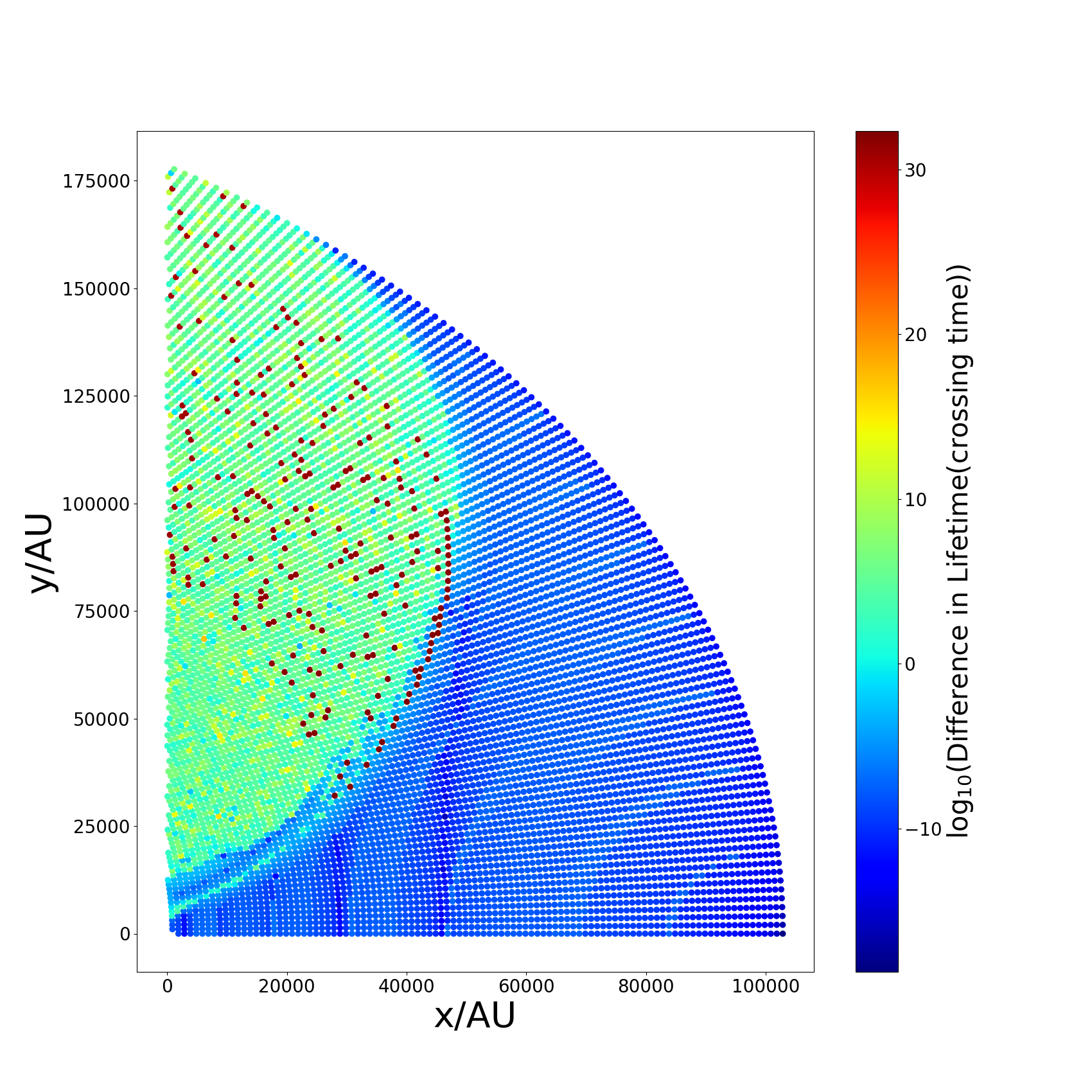}
    \caption{The log$_{10}$(lifetime/crossing time) for simulations in the Agekian Anosova region for Run Four of Set One simulations. See Table \ref{tab:SetOne}.}
    \label{fig:lifepostiveaxis}
\end{figure}

In the scenario where the spinning black hole is kept fixed on the positive axis, the log$_{10}$(lifetime/crossing time) of the systems appear Figure \ref{fig:lifepostiveaxis}. Longer living systems cluster to the top left -- away from the more massive rotating black hole, while systems that start off closer to it (bottom right) end very quickly. 

\begin{thebibliography}{00}


\bibitem[Abbott(2019)]{LIGO1}Abbott, B., Abbott, R., Abbott, T., Abraham, S., Acernese, F., Ackley, K., Adams, C., Adhikari, R., Adya, V., Affeldt, C., Agathos, M., Agatsuma, K., Aggarwal, N., Aguiar, O., Aiello, L., Ain, A., Ajith, P., Allen, G., Allocca, A., Aloy..., M. \& Zweizig, J. GWTC-1: A Gravitational-Wave Transient Catalog of Compact Binary Mergers Observed by LIGO and Virgo during the First and Second Observing Runs. {\em Phys. Rev. X}. \textbf{9}, 031040 (2019,9), https://link.aps.org/doi/10.1103/PhysRevX.9.031040
\bibitem[Abbott (2021)]{LIGO2}Abbott, R., Abbott, T., Abraham, S., Acernese, F., Ackley, K., Adams, A., Adams, C., Adhikari, R., Adya, V., Affeldt, C., Agathos, M., Agatsuma, K., Aggarwal, N., Aguiar, O., Aiello, L., Ain, A., Ajith, P., Akcay, S., Allen, G., Allocca..., A. \& Zweizig, J. GWTC-2: Compact Binary Coalescences Observed by LIGO and Virgo during the First Half of the Third Observing Run. {\em Phys. Rev. X}. \textbf{11}, 021053 (2021,6), https://link.aps.org/doi/10.1103/PhysRevX.11.021053
\bibitem[Chitan (2021)]{Chitan2021}Chitan, A., Mylläri, A. \& Haque, S. Relativistic effects on triple black holes: Burrau’s problem revisited. {\em Monthly Notices Of The Royal Astronomical Society}. \textbf{509}, 1919-1928 (2021,10), https://doi.org/10.1093/mnras/stab3124
\bibitem[Valtonen (1995)]{valtonen1995}Valtonen, M., Mikkola, S. \& Pietila, H. Burrau's three-body problem in the post-Newtonian approximation. {\em Monthly Notices Of The Royal Astronomical Society}. \textbf{273}, 751-754 (1995,4)
\bibitem[Peetre (1995)]{Peetre1995}Peetre, J. Outline of a scientific biography of Ernst Meissel (1826-1895). {\em Historia Mathematica}. \textbf{22}, 154-178 (1995,5), https://doi.org/10.1006
\bibitem[Valtonen (2016)]{valtonenetal2016}Valtonen, M., Anosova, J., Kholshevnikov, K., Mylläri, A., Orlov, V. \& Tanikawa, K. From Comets to Chaos. {\em The Three-body Problem From Pythagoras To Hawking}. pp. 51-84 (2016)
\bibitem[Burrau (1913)]{Burrau1913}Burrau, C. Numerische Berechnung eines Spezialfalles des Dreikörperproblems. {\em Astronomische Nachrichten}. \textbf{195}, 113-118 (1913), https://doi.org/10.1002
\bibitem[Rhook (2005)]{Rhook2005}Rhook, K. \& Wyithe, J. Realistic event rates for detection of supermassive black hole coalescence by LISA. {\em Monthly Notices Of The Royal Astronomical Society}. \textbf{361}, 1145-1152 (2005,8), https://doi.org/10.1111/j.1365-2966.2005.08987.x
\bibitem[McMillan (2001)]{Mcmillan2001}McMillan, S. \& Portegies Zwart, S. Black Hole Mergers in the Universe. {\em Dynamics Of Star Clusters And The Milky Way}. \textbf{228} pp. 517 (2001,1)
\bibitem[Belcyz (2002)]{Belcyz2002}Belczynski, K., Kalogera, V. \& Bulik, T. A Comprehensive Study of Binary Compact Objects as Gravitational Wave Sources: Evolutionary Channels, Rates, and Physical Properties. {\em The Astrophysical Journal}. \textbf{572}, 407-431 (2002,6)
\bibitem[Rodriguez (2018)]{Rodriguez2018}Rodriguez, C. \& Antonini, F. A Triple Origin for the Heavy and Low-spin Binary Black Holes Detected by LIGO/VIRGO. {\em The Astrophysical Journal}. \textbf{863}, 7 (2018,8), https://doi.org/10.3847/1538-4357/aacea4
\bibitem[Kormendy (1995)]{Kormendy1995}Kormendy, J. \& Richstone, D. Inward Bound—The Search for Supermassive Black Holes in Galactic Nuclei. {\em Annual Review Of Astronomy And Astrophysics}. \textbf{33}, 581-624 (1995,9), https://doi.org/10.1146
\bibitem[Valtonen (1996)]{Valtonen1996}Valtonen, M. Triple black hole systems formed in mergers of galaxies. {\em Monthly Notices Of The Royal Astronomical Society}. \textbf{278}, 186-190 (1996,1), https://doi.org/10.1093/mnras/278.1.186
\bibitem[Hoffman (2007)]{Hoffman2007}Hoffman, L. \& Loeb, A. Dynamics of triple black hole systems in hierarchically merging massive galaxies. {\em Monthly Notices Of The Royal Astronomical Society}. \textbf{377}, 957-976 (2007,5), https://doi.org/10.1111/j.1365-2966.2007.11694.x
\bibitem[Sigurdsson (1993)]{Sigurdsson1993}Sigurdsson, S. \& Hernquist, L. Primordial black holes in globular clusters. {\em Nature}. \textbf{364}, 423-425 (1993,7)
\bibitem[Vitral (2021)]{Vitral2021}Vitral, Eduardo \& Mamon, Gary A. Does NGC 6397 contain an intermediate-mass black hole or a more diffuse inner subcluster?. {\em A\&A}. \textbf{646} pp. A63 (2021), https://doi.org/10.1051/0004-6361/202039650
\bibitem[Kollatschny (2020)]{Kollatschny2020}Kollatschny, W., Weilbacher, P., Ochmann, M., Chelouche, D., Monreal-Ibero, A., Bacon, R. \& Contini, T. NGC 6240: A triple nucleus system in the advanced or final state of merging. {\em Astronomy \& Astrophysics}. \textbf{633} pp. A79 (2020,1), https://doi.org/10.1051
\bibitem[Yadav (2021)]{Yadav2021}Yadav, Jyoti, Das, Mousumi, Barway, Sudhanshu \& Combes, Francoise A triple active galactic nucleus in the NGC 7733-7734 merging group. {\em A\&A}. \textbf{651} pp. L9 (2021), https://doi.org/10.1051/0004-6361/202141210
\bibitem[Gomez (2021)]{Gomez2021}Vigna-Gómez, A., Toonen, S., Ramirez-Ruiz, E., Leigh, N., Riley, J. \& Haster, C. Massive Stellar Triples Leading to Sequential Binary Black Hole Mergers in the Field. {\em The Astrophysical Journal}. \textbf{907}, L19 (2021,1), https://doi.org/10.3847/2041-8213/abd5b7
\bibitem[Szebehely (1967)]{Szebehely1967}Szebehely, V. \& Peters, C. Complete solution of a general problem of three bodies. {\em The Astronomical Journal}. \textbf{72} pp. 876 (1967,9), https://doi.org/10.1086
\bibitem[Heggie (1996)]{Heggie1996}Heggie, D. \& Rasio, F. The Effect of Encounters on the Eccentricity of Binaries in Clusters. {\em Monthly Notices Of The Royal Astronomical Society}. \textbf{282}, 1064-1084 (1996,10)
\bibitem[Stone (2019)]{stone2019}Stone, N. \& Leigh, N. A statistical solution to the chaotic, non-hierarchical three-body problem. {\em Nature}. \textbf{576}, 406-410 (2019,12)
\bibitem[Hamers (2019)]{Hamers2019}Hamers, A. \& Samsing, J. Analytic computation of the secular effects of encounters on a binary: features arising from second-order perturbation theory. {\em Monthly Notices Of The Royal Astronomical Society}. \textbf{487}, 5630-5648 (2019,8)
\bibitem[Ginat (2021)]{ginat2021}Ginat, Y. \& Perets, H. Analytical, Statistical Approximate Solution of Dissipative and Nondissipative Binary-Single Stellar Encounters. {\em Physical Review X}. \textbf{11}, e031020 (2021,7)
\bibitem[Hut (1983a)]{Hut1}Hut, P. \& Bahcall, J. Binary-single star scattering. I - Numerical experiments for equal masses. {\em The Astrophysical Journal}. \textbf{268} pp. 319-341 (1983,5)
\bibitem[Hut (1983b)]{Hut2}Hut, P. Binary-single star scattering. II - Analytic approximations for high velocity. {\em The Astrophysical Journal}. \textbf{268} pp. 342-355 (1983,5)
\bibitem[Hut (1993a)]{Hut3}Hut, P. Binary–Single-Star Scattering. III. Numerical Experiments for Equal-Mass Hard Binaries. {\em The Astrophysical Journal}. \textbf{403} pp. 256 (1993,1)
\bibitem[Heggie (1993)]{Hut4}Heggie, D. \& Hut, P. Binary–Single-Star Scattering. IV. Analytic Approximations and Fitting Formulae for Cross Sections and Reaction Rates. {\em The Astrophysical Journals}. \textbf{85} pp. 347 (1993,4)
\bibitem[Goodman (1993)]{Hut5}Goodman, J. \& Hut, P. Binary–Single-Star Scattering. V. Steady State Binary Distribution in a Homogeneous Static Background of Single Stars. {\em The Astrophysical Journal}. \textbf{403} pp. 271 (1993,1)
\bibitem[McMillan (1996)]{Hut6}McMillan, S. \& Hut, P. Binary–Single-Star Scattering. VI. Automatic Determination of Interaction Cross Sections. {\em The Astrophysical Journal}. \textbf{467} pp. 348 (1996,8)
\bibitem[Heggie (1996)]{Hut7}Heggie, D., Hut, P. \& McMillan, S. Binary–Single-Star Scattering. VII. Hard Binary Exchange Cross Sections for Arbitrary Mass Ratios: Numerical Results and Semianalytic FITS. {\em The Astrophysical Journal}. \textbf{467} pp. 359 (1996,8)
\bibitem[Monaghan (1976)]{monaghan1976}Monaghan, J. A statistical theory of the disruption of three-body systems - I. Low angular momentum.. {\em Monthly Notices Of The Royal Astronomical Society}. \textbf{176} pp. 63-72 (1976,7)
\bibitem[Anosova (1990)]{Anosova1990}Anosova, J., Orlov, V., Chernin, A., Ivanov, A. \& Kiseleva, L. Dynamics and Configurations of Galaxy Triplets. {\em International Astronomical Union Colloquium}. \textbf{124} pp. 633-643 (1990), http://doi.org/10.1017/s0252921100005765
\bibitem[Anosova (1992)]{Anosova1992}Anosova, Z. \& Orlov, V. The types of motion in hierarchical and non-hierarchical triple systems - Numerical experiments. {\em Astronomy And Astrophysics}. \textbf{260}, 473-484 (1992,7)
\bibitem[Kerr (1963)]{Kerr1963}Kerr, R. Gravitational Field of a Spinning Mass as an Example of Algebraically Special Metrics. {\em Phys. Rev. Lett.}. \textbf{11}, 237-238 (1963,9), https://link.aps.org/doi/10.1103/PhysRevLett.11.237
\bibitem[Israel (1967)]{Israel1967}Israel, W. Event Horizons in Static Vacuum Space-Times. {\em Physical Review}. \textbf{164}, 1776-1779 (1967,12)
\bibitem[Israel (1968)]{Israel1968}Israel, W. Event horizons in static electrovac space-times. {\em Communications In Mathematical Physics}. \textbf{8}, 245-260 (1968,9)
\bibitem[Hawking (1972)]{Hawking1972}Hawking, S. Black holes in general relativity. {\em Communications In Mathematical Physics}. \textbf{25}, 152-166 (1972,6)
\bibitem[Lense (1918)]{Lense1918}Lense, J. \& Thirring, H. Über den Einfluß der Eigenrotation der Zentralkörper auf die Bewegung der Planeten und Monde nach der Einsteinschen Gravitationstheorie. {\em Physikalische Zeitschrift}. \textbf{19} pp. 156 (1918,1)
\bibitem[Ciufolini (2004)]{Ciufolini2004}Ciufolini, I. \& Pavlis, E. A confirmation of the general relativistic prediction of the Lense–Thirring effect. {\em Nature}. \textbf{431} pp. 958-960 (2004)
\bibitem[Fang (2019a)]{Fang1}Fang, Y. \& Huang, Q. Secular evolution of compact binaries revolving around a spinning massive black hole. {\em Phys. Rev. D}. \textbf{99}, 103005 (2019,5), https://link.aps.org/doi/10.1103/PhysRevD.99.103005
\bibitem[Fang (2019b)]{Fang2}Fang, Y., Chen, X. \& Huang, Q. Impact of a Spinning Supermassive Black Hole on the Orbit and Gravitational Waves of a Nearby Compact Binary. {\em The Astrophysical Journal}. \textbf{887}, 210 (2019,12), https://doi.org/10.3847/1538-4357/ab510e
\bibitem[Fang (2020)]{Fang3}Fang, Y. \& Huang, Q. Three body first post-Newtonian effects on the secular dynamics of a compact binary near a spinning supermassive black hole. {\em Phys. Rev. D}. \textbf{102}, 104002 (2020,11), https://link.aps.org/doi/10.1103/PhysRevD.102.104002
\bibitem[Mikkola (1993)]{Mikkola1993}Mikkola, S. \& Aarseth, S. An implementation ofN-body chain regularization. {\em Celestial Mechanics \& Dynamical Astronomy}. \textbf{57}, 439-459 (1993,11), https://doi.org/10.1007\$
\bibitem[Mikkola (1996)]{Mikkola1996}Mikkola, S. \& Aarseth, S. A slow-down treatment for close binaries. {\em Celestial Mechanics \& Dynamical Astronomy}. \textbf{64}, 197-208 (1996), https://doi.org/10.1007
\bibitem[Mikkola (2002)]{Mikkola2002}Mikkola, S. \& Aarseth, S. A Time-Transformed Leapfrog Scheme. {\em Celestial Mechanics And Dynamical Astronomy}. \textbf{84} pp. 343-354 (2002)
\bibitem[Hellstrom (2010)]{Hellstrom2010}Hellström, C. \& Mikkola, S. Explicit algorithmic regularization in the few-body problem for velocity-dependent perturbations. {\em Celestial Mechanics And Dynamical Astronomy}. \textbf{106}, 143-156 (2010,1), https://doi.org/10.1007
\bibitem[Mikkola (2006)]{Mikkola2006}Mikkola, S. \& Merritt, D. Algorithmic regularization with velocity-dependent forces. {\em Monthly Notices Of The Royal Astronomical Society}. \textbf{372}, 219-223 (2006,10), https://doi.org/10.1111
\bibitem[Mikkola (2008)]{Mikkola2008}Mikkola, S. \& Merritt, D. Implementing Few-Body Algorithmic Regularization with post-Newtonian Terms. {\em The Astronomical Journal}. \textbf{135}, 2398-2405 (2008,5), https://doi.org/10.1088
\bibitem[Mikkola (1999)]{Mikkola1999}Mikkola, S. \& Tanikawa, K. Explicit Symplectic Algorithms For Time‐Transformed Hamiltonians. {\em Celestial Mechanics And Dynamical Astronomy Volume }. \textbf{74} pp. 287-295 (1999)
\bibitem[Mikkola (2013)]{MikkolaElsev.2013}Mikkola, S. \& Tanikawa, K. Implementation of an efficient logarithmic-Hamiltonian three-body code. {\em New Astronomy}. \textbf{20} pp. 38-41 (2013,4), https://doi.org/10.1016
\bibitem[Mikkola (2013)]{Mikkola2013}Mikkola, S. \& Tanikawa, K. Regularizing dynamical problems with the symplectic logarithmic Hamiltonian leapfrog. {\em Monthly Notices Of The Royal Astronomical Society}. \textbf{430}, 2822-2827 (2013,2), https://doi.org/10.1093
\bibitem[Bardeen (1973)]{Bardeen1973}Bardeen, J., Carter, B. \& Hawking, S. The four laws of black hole mechanics. {\em Communications In Mathematical Physics}. \textbf{31}, 161-170 (1973,6)
\bibitem[Wald (1999)]{Wald1999}Wald, R. Gravitational Collapse and Cosmic Censorship. {\em Black Holes, Gravitational Radiation And The Universe: Essays In Honor Of C.V. Vishveshwara}. pp. 69-86 (1999), https://doi.org/10.1007/978-94-017-0934-7
\bibitem[Barker (1975)]{Barker1975}Barker, B. \& O'Connell, R. Gravitational two-body problem with arbitrary masses, spins, and quadrupole moments. {\em Physical Review D}. \textbf{12}, 329-335 (1975,7), https://doi.org/10.1103
\bibitem[Valtonen (2010)]{Valtonen2010}Valtonen, M., Mikkola, S., Merritt, D., Gopakumar, A., Lehto, H., Hyvönen, T., Rampadarath, H., Saunders, R., Basta, M. \& Hudec, R. MEASURING THE SPIN OF THE PRIMARY BLACK HOLE IN OJ287. {\em The Astrophysical Journal}. \textbf{709}, 725-732 (2010,1), https://doi.org/10.1088
\bibitem[Mikkola (2020)]{Mikkola2020}Mikkola, S. Motion in the Field of a Black Hole. {\em Gravitational Few-Body Dynamics}. pp. 211-220 (2020,4), https://doi.org/10.1017
\bibitem[Stella (2009)]{Stella2009}Stella, L. \& Possenti, A. Lense-Thirring Precession in the Astrophysical Context. {\em Space Science Reviews}. \textbf{148}, 105-121 (2009,12), https://doi.org/10.1007
\bibitem[Valtonen (2016)]{Valtonen2016}Valtonen, M., Zola, S., Ciprini, S., Gopakumar, A., Matsumoto, K., Sadakane, K., Kidger, M., Gazeas, K., Nilsson, K., Berdyugin, A., Piirola, V., Jermak, H., Baliyan, K., Alicavus, F., Boyd, D., Campas Torrent, M., Campos, F., Carrillo Gómez, J., Caton, D., Chavushyan, V., Dalessio, J., Debski, B., Dimitrov, D., Drozdz, M., Er, H., Erdem, A., Escartin Pérez, A., Fallah Ramazani, V., Filippenko, A., Ganesh, S., Garcia, F., Gómez Pinilla, F., Gopinathan, M., Haislip, J., Hudec, R., Hurst, G., Ivarsen, K., Jelinek, M., Joshi, A., Kagitani, M., Kaur, N., Keel, W., LaCluyze, A., Lee, B., Lindfors, E., Lozano de Haro, J., Moore, J., Mugrauer, M., Naves Nogues, R., Neely, A., Nelson, R., Ogloza, W., Okano, S., Pandey, J., Perri, M., Pihajoki, P., Poyner, G., Provencal, J., Pursimo, T., Raj, A., Reichart, D., Reinthal, R., Sadegi, S., Sakanoi, T., Salto González, J., Sameer, Schweyer, T., Siwak, M., Soldán Alfaro, F., Sonbas, E., Steele, I., Stocke, J., Strobl, J., Takalo, L., Tomov, T., Tremosa Espasa, L., Valdes, J., Valero Pérez, J., Verrecchia, F., Webb, J., Yoneda, M., Zejmo, M., Zheng, W., Telting, J., Saario, J., Reynolds, T., Kvammen, A., Gafton, E., Karjalainen, R., Harmanen, J. \& Blay, P. Primary Black Hole Spin in OJ 287 as Determined by the General Relativity Centenary Flare. {\em The Astrophysical Journal}. \textbf{819}, eL37 (2016,3)
\bibitem[Thompson (2019)]{Thompson2019}Thompson, T., Kochanek, C., Stanek, K., Badenes, C., Post, R., Jayasinghe, T., Latham, D., Bieryla, A., Esquerdo, G., Berlind, P., Calkins, M., Tayar, J., Lindegren, L., Johnson, J., Holoien, T., Auchettl, K. \& Covey, K. A noninteracting low-mass black hole–giant star binary system. {\em Science}. \textbf{366}, 637-640 (2019,11), https://doi.org/10.1126
\bibitem[Raidal (2019)]{Raidal2019}Raidal, M., Spethmann, C., Vaskonen, V. \& Veermäe, H. Formation and evolution of primordial black hole binaries in the early universe. {\em Journal Of Cosmology And Astroparticle Physics}. \textbf{2019}, 018-018 (2019,2), https://doi.org/10.1088/1475-7516/2019/02/018
\bibitem[Gonzalez (2007)]{Gonzalez2007}González, J., Hannam, M., Sperhake, U., Brügmann, B. \& Husa, S. Supermassive Recoil Velocities for Binary Black-Hole Mergers with Antialigned Spins. {\em American Physical Society}. \textbf{98}, e231101 (2007,6)
\bibitem[Campanelli (2007)]{Campanelli2007}Campanelli, M., Lousto, C., Zlochower, Y. \& Merritt, D. Large Merger Recoils and Spin Flips from Generic Black Hole Binaries.  (2007,4)
\bibitem[Karttunen (2006)]{karttunen}Valtonen, M. \& Karttunen, H. The Three-Body Problem. (Cambridge University Press,2006)
\bibitem[Chitan (2023)]{Chitan2023}Chitan, A., Myllari, A. \& Valtonen, M. Relativistic Effects on Triple Black Holes II The Influence of Spin in Burrau’s Problem. 1-3..  (2023,1), https://figshare.com/articles/dataset/Relativistic\_Effects\_on\_Triple\_Black\_Holes\_II\_The\_Influence\_of\_Spin\_in\_Burrau\_s\_Problem\_1-3\_/19735960
\bibitem[Will (2014b)]{Will2014b}Will, C. Post-Newtonian effects in N-body dynamics: conserved quantities in hierarchical triple systems. {\em Classical And Quantum Gravity}. \textbf{31}, 244001 (2014,12), https://doi.org/10.1088
\bibitem[Will (2014a)]{Will2014a}Will, C. Incorporating post-Newtonian effects in N-body dynamics. {\em Physical Review D}. \textbf{89} (2014,2), https://doi.org/10.1103
\bibitem[Mora (2004)]{MoraWill2004}Mora, T. \& Will, C. Post-Newtonian diagnostic of quasiequilibrium binary configurations of compact objects. {\em Phys. Rev. D}. \textbf{69}, 104021 (2004,5), https://link.aps.org/doi/10.1103/PhysRevD.69.104021
\bibitem[Mikkola (2020)]{mikkolabook2020}Mikkola, S. Motion in the Field of a Black Hole. {\em Gravitational Few-Body Dynamics}. pp. 211-220 (2020,4), https://doi.org/10.1017
\bibitem[Kidder (1995)]{kidder1995}Kidder, L. Coalescing binary systems of compact objects to (post)<sup>5/2</sup>-Newtonian order. V. Spin effects. {\em Phys. Rev. D}. \textbf{52}, 821-847 (1995,7), https://link.aps.org/doi/10.1103/PhysRevD.52.821
\bibitem[Anosova (1994)]{anosova1994}Anosova, J., Orlov, V. \& Aarseth, S. Initial Conditions and Dynamics of Triple Systems. {\em Celestial Mechanics And Dynamical Astronomy}. \textbf{60}, 365-372 (1994,11)
\bibitem[Anosova (1991)]{anosova1991}Anosova, Z. \& Nebukin, A. On the representativity of initial conditions of triple systems. {\em Astronomy And Astrophysics}. \textbf{252}, 410-413 (1991,12)
\bibitem[Anosova (1984)]{anosova1984}Anosova, Z. \& Orlov, V. The initial configuration and the escape of triple systems with components of different masses. {\em Trudy Astronomicheskoj Observatorii Leningrad}. \textbf{39}, 101-111 (1984,1)
\bibitem[Boekholt (2021)]{boekholt2021}Boekholt, T., Moerman, A. \& Portegies Zwart, S. Relativistic Pythagorean three-body problem. {\em Phys. Rev. D}. \textbf{104}, 083020 (2021,10), https://link.aps.org/doi/10.1103/PhysRevD.104.083020
\bibitem[Peißker (2024)]{imbh2024}Peißker, F., Zajaček, M., Labaj, M., Thomkins, L., Elbe, A., Eckart, A., Labadie, L., Karas, V., Sabha, N., Steiniger, L. \& Melamed, M. The Evaporating Massive Embedded Stellar Cluster IRS 13 Close to Sgr A*. II. Kinematic Structure. {\em The Astrophysical Journal}. \textbf{970}, e74 (2024,7)
\bibitem[Lam (2022)]{lam2022}Lam, C., Lu, J., Udalski, A., Bond, I., Bennett, D., Skowron, J., Mróz, P., Poleski, R., Sumi, T., Szymański, M., Kozłowski, S., Pietrukowicz, P., Soszyński, I., Ulaczyk, K., Wyrzykowski, Ł., Miyazaki, S., Suzuki, D., Koshimoto, N., Rattenbury, N., Hosek, M., Abe, F., Barry, R., Bhattacharya, A., Fukui, A., Fujii, H., Hirao, Y., Itow, Y., Kirikawa, R., Kondo, I., Matsubara, Y., Matsumoto, S., Muraki, Y., Olmschenk, G., Ranc, C., Okamura, A., Satoh, Y., Silva, S., Toda, T., Tristram, P., Vandorou, A., Yama, H., Abrams, N., Agarwal, S., Rose, S. \& Terry, S. An Isolated Mass-gap Black Hole or Neutron Star Detected with Astrometric Microlensing. {\em The Astrophysical Journal}. \textbf{933}, eL23 (2022,7)
\bibitem[Sahu (2022)]{sahu2022}Sahu, K., Anderson, J., Casertano, S., Bond, H., Udalski, A., Dominik, M., Calamida, A., Bellini, A., Brown, T., Rejkuba, M., Bajaj, V., Kains, N., Ferguson, H., Fryer, C., Yock, P., Mróz, P., Kozłowski, S., Pietrukowicz, P., Poleski, R., Skowron, J., Soszyński, I., Szymański, M., Ulaczyk, K., Wyrzykowski, Ł., Barry, R., Bennett, D., Bond, I., Hirao, Y., Silva, S., Kondo, I., Koshimoto, N., Ranc, C., Rattenbury, N., Sumi, T., Suzuki, D., Tristram, P., Vandorou, A., Beaulieu, J., Marquette, J., Cole, A., Fouqué, P., Hill, K., Dieters, S., Coutures, C., Dominis-Prester, D., Bennett, C., Bachelet, E., Menzies, J., Albrow, M., Pollard, K., Gould, A., Yee, J., Allen, W., Almeida, L., Christie, G., Drummond, J., Gal-Yam, A., Gorbikov, E., Jablonski, F., Lee, C., Maoz, D., Manulis, I., McCormick, J., Natusch, T., Pogge, R., Shvartzvald, Y., Jørgensen, U., Alsubai, K., Andersen, M., Bozza, V., Novati, S., Burgdorf, M., Hinse, T., Hundertmark, M., Husser, T., Kerins, E., Longa-Peña, P., Mancini, L., Penny, M., Rahvar, S., Ricci, D., Sajadian, S., Skottfelt, J., Snodgrass, C., Southworth, J., Tregloan-Reed, J., Wambsganss, J., Wertz, O., Tsapras, Y., Street, R., Bramich, D., Horne, K., Steele, I. \& RoboNet Collaboration An Isolated Stellar-mass Black Hole Detected through Astrometric Microlensing. {\em The Astrophysical Journal}. \textbf{933}, e83 (2022,7)
\bibitem[Lam (2023)]{Lam2023}Lam, C. \& Lu, J. A Reanalysis of the Isolated Black Hole Candidate OGLE-2011-BLG-0462/MOA-2011-BLG-191. {\em The Astrophysical Journal}. \textbf{955}, e116 (2023,10)


\end{thebibliography}




\end{document}